\documentclass{aa}

\usepackage{graphicx}
\usepackage{natbib}
\bibpunct{(}{)}{;}{a}{}{,}
\usepackage{txfonts}
\usepackage{float}
\usepackage{multirow}
\usepackage{tikz,tikz-3dplot}\usetikzlibrary{calc}
\usepackage{mathtools}
\usepackage[makeroom]{cancel}
\usepackage{caption}
\usepackage{subcaption}
\usepackage{xcolor}
\usepackage{hyperref}
\hypersetup{colorlinks,breaklinks,
  linkcolor=[rgb]{0.368417, 0.506779, 0.709798},
  citecolor=[rgb]{0.368417, 0.506779, 0.709798},
  urlcolor=[rgb]{0.368417, 0.506779, 0.709798}}

\usepackage{silence}
\usepackage{siunitx}
\newcommand{\mm}{\textcolor{magenta}}

\begin{document}

   \title{Post-Newtonian orbital mechanics around a black hole in modified gravity}

   \author{I. Liodis
          \inst{1,2,3,4,5}\fnmsep\thanks{Corresponding authors: I.~Liodis (ioannis.liodis@desy.de)
          .
          }
          \and
          G. Hei{\ss}el\inst{3}
          \and
          R. Mastroianni\inst{3}
          \and
          J. Grover\inst{3}
          \and
          D. Izzo\inst{3}
          }
\institute{
Deutsches Elektronen-Synchrotron DESY, Platanenallee 6, 15738 Zeuthen, Germany
\and {Institute of Physics and Astronomy, University of Potsdam, 28, Karl-Liebknecht-Straße 24/25, 14476 Potsdam, Germany}
\and {Advanced Concepts Team, European Space Agency, TEC-SF, ESTEC, Keplerlaan 1, 2201 AZ Noordwijk, The Netherlands}
\and {Deutsche Zentrum für Astrophysik DZA, Postplatz 1, 02826 Görlitz, Germany}
\and {Gravitation and Astroparticle Physics Amsterdam (GRAPPA), University of Amsterdam, Science Park 904, 1098 XH Amsterdam, The Netherlands}
             }

  \abstract
   {Scalar-tensor-vector gravity, also known as modified gravity (MOG), has emerged as an alternative to General Relativity (GR). It aims to explain astrophysical phenomena without invoking dark matter. The theory introduces a dynamic scalar field, a vector field, and modifications to the gravitational constant. The S-stars orbiting the supermassive black hole at the Galactic centre provide a unique opportunity to test the predictions of MOG because the orbital measurements are highly precise.}
   {We investigate the perturbations in the orbits of S-stars under MOG, focusing on the effects on orbital elements, observables such as right ascension, declination, and radial velocity, and the potential degeneracy with dark matter scenarios.}
   {We numerically integrated the first post-Newtonian equations of motion for S-stars within the MOG framework, considering contributions from the space-time geometry and the fifth force. We analysed the time evolution of orbital elements and projected the orbits onto the plane of the sky to assess deviations from GR. Furthermore, we compared the MOG-induced effects with those expected from a dark matter distribution.}
   {We found that MOG significantly alters the orbital precession, particularly for higher values of the MOG parameter $\alpha$. For sufficiently large $\alpha$ or long observational baselines, the deviations in the observables can reach amplitudes comparable to current observational precision. Furthermore, we demonstrate that MOG effects can mimic those of a dark matter distribution, particularly in the argument of pericentre, and we reveal an unexplored connection between MOG and GR with electromagnetism.}
   {The effects of MOG on stellar orbits are distinct from those predicted by GR and can be tested with precise astrometric and spectroscopic measurements of the S-stars. However, a potential degeneracy with dark matter signatures necessitates careful interpretation of observational data. Further observations over longer periods are required to conclusively distinguish between these scenarios.}


    \maketitle

\section{Introduction}\label{S:introduction}
High-precision measurements of the stellar dynamics in the Galactic centre have allowed for direct tests of modified theories of gravity \citep{Borka_2013,Della_Monica_2021,PhysRevD.99.044053,2019PhRvD..99b4052R, Zakharov_2018} and dark mass (DM) (DM includes dark matter and any other faint source \citep{Heissel+22,Lechien_2024,Yuan_2022}). Namely, the contribution of Schwarzschild and mass precessions on the orbit of S2 around Sgr~A* has been studied \citep{Heissel+22}, and a method for constraining the DM was set up assuming two different density profiles around the supermassive black hole. A more general density profile has been used as well to reconstruct DM \citep{Lechien_2024}, which requires significantly more observations to achieve the same level of accuracy in setting bounds on the mass, however.

Constraints on the parametrised post-Newtonian (PPN) parameters $\beta_{PPN}$ and $\gamma_{PPN}$ were also imposed using the orbits of S-stars \citep{Gainutdinov_2020}, and constraints were forecast for future S2 and S62 observations \citep{de_Mora_Losada_2025}. The PPN formalism is independent of a particular modified theory because it provides a general framework that parametrises deviations from GR. When strong constraints are to be placed on a particular theory of gravity, however, PPN is not optimal. As a more general framework, it considers more parameters than in the case of a particular theory. For instance, based on the orbital precession of the S2 star in particular, scalar-tensor-vector gravity (STVG; see below) has been subject to tests \citep{Della_Monica_2021} under a fully relativistic framework. Even though the latter is formally complete, it is not necessarily the most intuitive and handy approach for investigating effects of a modified theory. When far enough from the black hole, as is the case for the currently known S-stars, the post-Newtonian framework offers a greater flexibility, such as an easier way to add perturbative accelerations, and the ability to investigate the impact of these perturbations, as well as of other effects, in isolation.

In an attempt to construct a modified theory of gravity that can explain the missing mass in the galaxies without dark matter, John Moffat constructed the STVG, also known as modified gravity (MOG) \citep{Moffat_2006}. In this theory, the gravitational constant $G$, a vector field coupling constant $\omega$ as well as its mass $\mu_\phi$, are dynamical scalar fields that can vary in space and time. It should be noted that unlike modified Newtonian dynamics (MOND) \citep{1983ApJ...270..365M}, MOG theory is relativistic by construction. It can therefore be tested in a wide range of astrophysical and cosmological environments. Namely, MOG passes the wide binary test \citep{Moffat_2024} and accounts for galactic dynamics without postulating dark matter \citep{Moffat_2013,Brownstein_2006,Moffat_2014,Davari_2020,10.1093/mnras/stad3400,GREEN2019100323,Moffat_2015}. It can also explain the Bullet cluster \citep{Israel_2018,Brownstein_2007}, reproduces lensing solely with baryon mass \citep{PhysRevD.103.044045}, and fits or potentially explains cosmological data \citep{Moffat:2026ken,moffat2026stvgmogclusterdynamicscosmological,Davari_2021,2020EPJC...80..906M}. Furthermore, two novel observational signatures have recently been investigated in MOG black hole shadows perturbed by passing gravitational waves \citep{Lobos:2026jyp}. We focus on a particle orbiting a MOG black hole (BH). To do this, the space-time geometry around compact objects is described in the next paragraph.

\section{Background: Modified gravity, and the MOG black hole}\label{sec:background}
Following \cite{Moffat_2021}, we review the theoretical basis of MOG in Sect.~\ref{sec:stvg}, which is a truncated version of the initial theory proposed by \cite{Moffat_2006}. In Sect.~\ref{sec:tp_motion_fifth_force_and_classical_limit} we briefly review the equations governing test particle motion, the fifth force, and the classical limit of the theory. We then focus on the case of a MOG black hole in Sect.~\ref{sec:mog bh}, which is studied in more detail throughout the rest of this work. In the current paper, the metric signature convention $(+,-,-,-)$ is used.

\subsection{Modified gravity}
\label{sec:stvg}
    In General Relativity, the gravitational field is carried by the metric tensor $g_{\mu\nu}$. In addition to this symmetric field component, MOG adds two further aspects to the gravitational interaction: a massive vector field $\phi_\mu$, and a scalar field $\chi= 1/G$, where $G$ is the coupling strength of gravity. With $c=1$ units, the action is expressed as
    \begin{equation}
    S=S_G+S_\phi+S_M,
    \end{equation}
    where
    \begin{subequations}
    \begin{gather}
        \label{Baction_1}
        S_G=\frac{1}{16\pi}\int d^4x\sqrt{-g}\biggl(\chi R+\frac{\omega_M}{\chi}\nabla^\mu\chi\nabla_\mu\chi +2\Lambda\biggr)\,,\\
        \label{Baction}
        S_\phi=\int d^4x\sqrt{-g}\biggl(-\frac{1}{4}B^{\mu\nu}B_{\mu\nu}+\frac{1}{2}\mu_\phi^2\phi^\mu\phi_\mu\biggr)\, ,
    \end{gather}
    \end{subequations} 
    with $g=\det g_{\mu\nu}\,$, $R$ the Ricci scalar,
    $\Lambda$ the cosmological constant, $S_M$ the matter action, $B_{\mu\nu} \equiv \partial_\mu \phi_\nu - \partial_\nu \phi_\mu$ an anti-symmetric Faraday-like tensor, and $\omega_M$ a constant. Moreover, $G$ is expanded by $G=G_N(1+\alpha)$, and $\mu_\phi$ is the effective running mass of the spin $1$ graviton vector field. $G_N$ denotes Newton’s gravitational constant.

    To obtain the field equations, the action with respect to $g_{\mu\nu}$, $\phi_\mu$ and $\chi$ has to be varied, which leads to
    \begin{subequations}\label{eq:MOG_field_eqs}
    \begin{gather}
            \begin{split}
            \label{Gequation}
            G_{\mu\nu}=&-\frac{\omega_M}{\chi^2}\biggl(\nabla_\mu\chi\nabla_\nu\chi -\frac{1}{2}g_{\mu\nu}\nabla^\alpha\chi\nabla_\alpha\chi\biggr)\\
            &-\frac{1}{\chi}(\nabla_\mu\chi\nabla_\nu\chi-g_{\mu\nu}\Box\chi)+\frac{8\pi}{\chi}T_{\mu\nu},\end{split}\\
            \label{Bequation}
            \nabla_\nu B^{\mu\nu}+\mu_\phi^2\phi^\mu=J^\mu,\\
            \label{Boxchi}
            \Box\chi=\frac{8\pi}{(2\omega_M+3)}T,
    \end{gather}
    \end{subequations}
    respectively, where $G_{\mu\nu}$ is the usual Einstein tensor, $\Box=\nabla^\mu\nabla_\mu$, $J_\mu$ is the current matter source of the vector field, and finally, the energy-momentum tensor is
    \begin{equation}
    T_{\mu\nu}=T^M_{\mu\nu}+T^\phi_{\mu\nu}+
    g_{\mu\nu}\frac{\chi\Lambda}{8\pi}.
    \end{equation}
    The $T^\phi_{\mu\nu}$ term is defined as
    \begin{equation}
    T^\phi_{\mu\nu}=-\biggl({B_\mu}^\alpha B_{\alpha\nu}-\frac{1}{4}g_{\mu\nu}B^{\alpha\beta}B_{\alpha\beta}+\mu_\phi^2\phi_\mu\phi_\nu-\frac{1}{2}g_{\mu\nu}\phi^\alpha\phi_\alpha\biggr),
    \end{equation}
    and $T$ is the conventional trace $T=g^{\mu\nu}T_{\mu\nu}$.

    \subsection{Test-particle motion, the fifth force, and the classical limit}\label{sec:tp_motion_fifth_force_and_classical_limit}
    
    The action of a test-particle with mass $m$ is
    \begin{equation}\label{eq:MOG_tp_initial_action_natunits}
        S_{TP} = -m {c^2} \int d\tau - \frac{q}{{c}} \int \phi_\mu \frac{dx^\mu}{d \tau} d\tau,
    \end{equation}
    where the second term is added due to the fifth force ($\mathcal{FF}$) \citep{Moffat_2006}. This additional interaction is only relevant for massive particles, where $q = \sqrt{\alpha G_N} m $ is the particle $\mathcal{FF}$ charge, and $\phi_\mu$ is the vector field. The original scalar field is the inverse gravitational coupling strength $\chi = G^{-1} = (1+\alpha)^{-1}G_N^{-1}$, which for a MOG black hole is constant \citep{Moffat_2021}. We can therefore treat $\alpha$ as a dimensionless constant. In a general space-time characterised by Christoffel symbols $\Gamma^{\mu }_{\alpha \beta }$, the equation of motion of a test particle with four-velocity $U^\mu$ is obtained from the action \eqref{eq:MOG_tp_initial_action_natunits} \citep{Moffat_2015_BH}    \begin{equation}\label{eq:MOG_compact_eom}
        \dot{U} ^\mu +\Gamma ^{\mu }_{\alpha \beta }U^{\alpha } U^{\beta }=
        \begin{cases}
            0 & \text{massless test particle}\\
            \frac{q}{m}B^{\mu }_{\phantom{a} \nu }U^{\nu} & \text{massive test particle}
        \end{cases}.
    \end{equation}
    Due to the $\mathcal{FF}$ term on the right-hand side, the motion is non-geodesic in the massive case.
    
    As derived by \cite{Moffat_2006}, the acceleration of a massive particle in the weak-field regime due to a central point mass $M$ is given by
    \begin{equation}\label{eq:MOG_weak_limit}
        a_{\text{MOG}}(r) = -\frac{G_N M}{r^2} \left[
            1 + \alpha - \alpha e^{-\mu_\phi r}(1 + \mu_\phi r) \right].
    \end{equation}
    At large distances, when $\mu_\phi r \gg1$, the term proportional to the exponential function vanishes, and what survives is the Newtonian gravitational acceleration, but enhanced by a factor $(1+\alpha)$. Via this feature, the theory is able to describe the observed flatness of galaxy rotation curves without postulating additional mass \citep{Moffat_2013,Brownstein_2006,Davari_2020,10.1093/mnras/stad3400,GREEN2019100323,Moffat_2015}. When $\mu_\phi r \ll 1$, we can linearise the exponential term, resulting in a cancellation of $\alpha$, and recovering the Newtonian classical limit. This means that for small length scales (but still in the weak field), the repulsive vector field exactly cancels out the enhancement of the gravitational constant.

    \subsection{MOG black holes}
    \label{sec:mog bh}
    When we consider a BH in this theory, additional simplifications can be introduced. Namely, \cite{Moffat_2021} demonstrated that starting from the field equations \eqref{eq:MOG_field_eqs} for the matter-free $\phi_\mu$ field-vacuum case with zero cosmological constant, we can further set the running mass to zero (i.e. $\mu_\phi = 0$), which causes $\chi$ to be just a constant. The latter indicates that $\alpha$ from now on is a parameter that quantifies the fractional increase of the MOG gravitational constant $G$ compared to the Newtonian one $G_N$. \cite{Moffat_2009} also justified why the same simplifications can be admitted starting from the full general version of the theory.

    The static spherically symmetric solution can then be written as \citep{Moffat_2015_BH}
    \begin{equation}\label{eq:MOG_line_element}
        ds^2 = \Psi(r) c^2dt^2 -\Psi^{-1}(r) dr^2 - r^2d\Omega ^2,
    \end{equation}
    where $d \Omega^2 = d \theta^2 + \sin^2 \theta d \phi^2$ and
    \begin{equation}\label{eq:psi_MOG_1}
        \Psi(r) = 1-\frac{2GM}{c^2 r}+\frac{G Q^2}{c^4 r^2}.
    \end{equation}
    $M$ is the BH mass, and $Q = \sqrt{\alpha G_N} M$ denotes its $\mathcal{FF}$ charge. Setting $\alpha = 0$ recovers the Schwarzschild solution in GR. In Sect.~\ref{sec:reiss_nord} we comment on the relation between the line element in Eq.~\eqref{eq:MOG_line_element} and the Reissner-Nordström solution for a charged BH, as well as for the motion of test particles in both theories and space-times.
    
    Moreover, the vector field for a MOG black hole with mass $M$ in Schwarzschild-like coordinates is given by \citep{Moffat_2021,Della_Monica_2021}
    \begin{equation}\label{eq:phu_mu_sign}
        \phi_\mu = \left(\frac{\sqrt{\alpha G_N} M}{r},0,0,0 \right),
    \end{equation}
    where the choice of the sign is explained in Appendix \ref{appendix:ffandem}.

\section{Relation to Reissner-Nordström space-time}
\label{sec:reiss_nord}
        In GR, the space-time of a BH with mass $\Tilde{M}$ and electric charge $\Tilde{Q}$ is given by the Reissner-Nordström (RN) solution \citep{Reissner_1916,1918KNAB...20.1238N}. The line element in standard spherical coordinates and in Gaussian units (so that $1/(4\pi\varepsilon_0) = 1$) has the same form as Eq.~\eqref{eq:MOG_line_element}, with
        \begin{equation}
            \Psi(r)\mapsto\Psi_{\text{RN}}(r) = 1-\frac{2 G_N \Tilde{M}}{c^2 r}+\frac{G_N \Tilde{Q}^2}{c^4 r^2}.
        \end{equation}
        This relation with a MOG BH was pointed out \citep{Moffat_2009,Moffat_2015_BH} under the straightforward relabelling
        \begin{align}\label{eq:straight forward dict}
        G \mapsto G_N \quad\text{and}\quad Q \mapsto \Tilde{Q}.
        \end{align}
        
        Here, we present a different substitution, which clearly shows that not only are there striking similarities, but the two theories produce indistinguishable geometries and equations of motions under a certain fine-tuning of their parameters.
        In particular, writing Eq.~\eqref{eq:psi_MOG_1} as
        \begin{equation}
            \Psi_{\text{MOG}}(r) = 1-\frac{2 G_N \left[(1+\alpha) M\right]}{c^2 r}+\frac{G_N  \left[\sqrt{1+\alpha}Q\right]^2}{c^4 r^2},
        \end{equation}
        and identifying
        \begin{equation}\label{eq:tilde_identification}
            \Tilde{M} = (1+\alpha) M \quad \text{and} \quad \Tilde{Q} = \sqrt{1+\alpha} Q,
        \end{equation}
        leads to
        \begin{equation}\label{eq:Psi_identification}
            \Psi_{\text{MOG}}(r;M,Q) = \Psi_{\text{RN}}(r;\Tilde{M},\Tilde{Q}).
        \end{equation}
        In other words, under the fine-tuning of Eq.~\eqref{eq:tilde_identification}, the line elements of the two theories are indistinguishable, and hence, so are the left hand sides (geodesic part) of their equations of motion (expressed by Eq.~\eqref{eq:MOG_compact_eom}). With the above straightforward dictionary of Eq.~\eqref{eq:straight forward dict}, this equivalence is not achieved because the value of $\alpha$ affects $G$ and $Q$ both.

        Furthermore, the right-hand sides of the equations of motion (Eq.~\eqref{eq:MOG_compact_eom}) for massive test particles become indistinguishable for the two theories when the fine-tuning of Eq.~\eqref{eq:tilde_identification} is also applied to the test particle,
        \begin{align}
            \Tilde{m} = (1+\alpha) m \quad \text{and} \quad \Tilde{q} = \sqrt{1+\alpha} q.
        \end{align}
        In the RN case, this right-hand side term for a massive ($\Tilde{m}$) and electrically charged ($\Tilde{q}$) test particle is
        \begin{equation}\label{eq:tilde_identification_test_particle}
            \text{RHS}_{\text{RN}} = \frac{\Tilde{q}}{\Tilde{m}} {g^{\mu\rho}} (\partial_\rho A_\nu - \partial_\nu A_\rho) U^{\nu} \, , \text{ with }A_0 = \frac{\Tilde{Q}}{r}
        \end{equation}
        being the only non-zero component of the four potential. In the MOG case, the right-hand side for a test particle with mass $m$ and $\mathcal{FF}$ charge $q$ is
        \begin{equation}
            \text{RHS}_{\text{MOG}} = \frac{q}{m}  {g^{\mu\rho}}(\partial_\rho \phi_\nu - \partial_\nu \phi_\rho) U^{\nu}\,,\text{ with } \phi_0 = \frac{Q}{r}
        \end{equation}
        being the only non-zero component of the four potential. Imposing the fine-tuning of Eqs.~\eqref{eq:tilde_identification} and~\eqref{eq:tilde_identification_test_particle} then yields the equivalence of the two right-hand sides.

        We summarise some of the observations and implications of the above discussion below.
        \begin{itemize}
            \item Under the fine-tuning of Eq.~\eqref{eq:tilde_identification}, geodesics around a MOG BH with mass and fifth-force charge $(M,Q)$ are indistinguishable from the geodesics around a GR BH $(\Tilde{M},\Tilde{Q})$. Hence, massless test particles follow the same dynamics around a BH in both universes.
            \item Massive but electrically neutral test particles also follow the geodesics of the Reissner-Nordström space-time, while massive test particles around a MOG BH deviate from these geodesics due to the proportionality of the fifth-force charge $q$ to the mass. In other words, there is no MOG analogue to the electrically neutral massive test particle of GR.
            \item The additional fine-tuning of Eq.~\eqref{eq:tilde_identification_test_particle} also establishes an equivalence between the (non-geodesic) dynamics of massive particles around a MOG BH $(M,Q)$ and a charged GR BH $(\Tilde{M},\Tilde{Q})$.
        \end{itemize}

        It should be emphasised, however, that the MOG theory around a BH corresponds to a reduced version of the full theory, as discussed in Sec.~\ref{sec:mog bh}. Consequently, our fine-tuning procedure may not remain directly applicable in other systems, particularly when $\alpha$ is not constant.

        After establishing the above correspondence between the two theories, we demonstrate its utility through a brief discussion of the horizons of a MOG BH. The Reissner-Nordström BH horizons are given by
        \begin{equation}
            r_{\pm} = \frac{G_N \Tilde{M}}{c^2} \pm \frac{\sqrt{G_N^2 \Tilde{M}^2 - G_N \Tilde{Q}^2}}{c^2},
        \end{equation}
        and since any $\mathcal{FF}$ charge is given by $Q = \sqrt{\alpha G_N} M$, starting from~\eqref{eq:tilde_identification}, for a positive $\alpha$, we have
        \begin{equation}
            \frac{G_N \Tilde{M}}{\sqrt{G_N}\Tilde{Q}} = \sqrt{\frac{1+\alpha}{\alpha}}>1,
        \end{equation}
        meaning that a MOG BH always has two horizons \citep{Moffat_2015_BH}, and it can only approach the extremal case when $\alpha \rightarrow \infty$.

        In conclusion, a MOG BH is a very specific case of the RN solution. The latter is free to take any mass $\Tilde{M}$ and electric charge $\Tilde{Q}$ (the same holds for every other massive object), meaning that on the mass-charge plane, every point is allowed. In contrast, a MOG BH (and again, every other massive object) will live on the straight line $\Tilde{Q} = \sqrt{\alpha G_N/(1+\alpha)} \Tilde{M}$ on this plane. Thus, by observing trajectories around BHs, we would not be able to infer whether our universe follows MOG or Einstein-Maxwell under the fine-tuning \eqref{eq:tilde_identification}, but now for every mass and charge. An alternative interpretation might be that a dark sector is associated with dark photons and charges, leading to a modified RN space-time \citep{MORRIS2023138325}. A compact object then has a total mass $M_{\text{tot}} = M_D + M_S$ and a total charge $Q_{\text{tot}} = Q_D + Q_S$, where the $D$ and $S$ indices refer to the dark and standard model charge, respectively. In this context, we can set $Q_S = 0$ and identify $M_S = M$, $M_D = \alpha M$ and $Q_D = \sqrt{1 + \alpha} Q$, such that $M_{\text{tot}} = \Tilde{M}$ and $Q_{\text{tot}} = \Tilde{Q}$, leading to indistinguishable dynamics once again.

\section{Post-Newtonian dynamics of a massive test particles orbiting a MOG black hole}\label{sec:PN dynamics}

    \subsection{Derivation of the equation of motion}
    \label{sec:eom}
        From now on, we recover the powers of $c$, such that the massive test-particle action of Eq.~\eqref{eq:MOG_tp_initial_action} is
        \begin{equation}\label{eq:MOG_tp_initial_action}
            S_{TP} = -m c^2 \int d\tau - \frac{q}{c} \int \phi_\mu \frac{dx^\mu}{d \tau} d\tau.
        \end{equation}
        The relative normalisation of powers of $c$ for the two components, as well as the relative signs, are consistent with a charged particle in an electromagnetic field \citep{Jackson:1998nia}. For the test particle orbiting a MOG black hole with mass $M$ (see Sect.~\ref{sec:mog bh}), the action $S_{TP}$ in Schwarzschild-like coordinates can be written as
        \begin{equation}\label{eq:tp_action}
            S_{TP} = \int dt \, L_{TP} = - \int dt \, m c^2 \left[ \frac{d \tau}{dt}  + \alpha \frac{G_N M}{r c^2} \right],
        \end{equation}
        where $L_{TP}$ denotes the test-particle Lagrangian. Our goal is to derive the equations of motion up to first post-Newtonian (1PN) order from this action. Following the method of \cite{Gainutdinov_2020}, we achieved this by first transforming to isotropic coordinates, then developing $L_{TP}$ to the desired powers of $c^{-2}$, and finally, transforming to Cartesian-like coordinates. The full steps can be found in Appendix~\ref{sec:1PN_derivation}. The Lagrangian, obtained by combining Eq.~\eqref{eq:lag1} and Eq.~\eqref{eq:FF}, is  
        \begin{align}        \begin{split}\label{eq:mog_lagrangian}
            \frac{L_{TP}}{m} =&
            \frac{\mathbf{\Dot{x}}^2}{2}
            \left[ 1 + \frac{\mathbf{\Dot{x}}^2}{4 c^2} - 3(\alpha \mathcal{M} +1) \frac{\varphi_N}{c^2}  \right] \\
            &- (\alpha \mathcal{M} + 1)\varphi_N \left[ 1 + (2 \alpha \mathcal{M} + 1)\frac{\varphi_N}{2 c^2} \right] \\
            &+ \mathcal{Q}\left\{
            \alpha \varphi_N + \alpha(1+\alpha)\frac{\varphi_N^2}{c^2} \right\}+
            \mathcal{O}(c^{-4}),
        \end{split}
        \end{align}
        where $\mathbf x$ denotes the spatial position vector in Cartesian-like coordinates, and $\varphi_N = - G_N M / |\mathbf{x}|$ denotes the Newtonian central mass potential. The two parameters $\mathcal{M}$, $ \mathcal{Q}$ were inserted by hand as book-keeping of the origin of the various terms. In the full theory, they have a numerical value equal to $1$. However, in Sect.~\ref{sec:MOG_impacts}, we occasionally also set them to zero in order to investigate each contribution in isolation. Terms not proportional to either book-keeping parameter are due to the Schwarzschild metric, and are thus also present in the 1PN expansion of GR. These are indeed the only surviving terms for $\alpha=0$, as expected. Terms proportional to $\mathcal{M}$ are due to the deviation of the MOG BH metric from the Schwarzschild metric when $\alpha>0$. Terms proportional to $\mathcal{Q}$ are due to the fifth force.
        
        Finally, we plugged the Lagrangian of Eq. \eqref{eq:mog_lagrangian} into the Euler-Lagrange equations to obtain the MOG equation of motion up to 1PN order (see also \citet{Gainutdinov_2020} for a related but less explicit result),
        \begin{align}
        \begin{split}\label{eq:MOG_eom}
            \ddot{\mathbf{x}} =
            & - \nabla \varphi_N -  \alpha \cancelto{\scriptstyle{0}}{(\mathcal{M} - \mathcal{Q})} \nabla \varphi_N \qquad\qquad\qquad\qquad\quad\;\text{(Kepler)} \\
            & - \nabla \varphi_N \left( 4\frac{\varphi_N}{c^2} + \frac{\Dot{\mathbf{x}}^2}{c^2} \right)
            + 4 \left(\nabla \varphi_N \cdot \frac{\Dot{\mathbf{x}}}{c}\right)\frac{\Dot{\mathbf{x}}}{c} \qquad\qquad\quad\text{(GR 1PN)} \\ 
            &+ \alpha (4\mathcal{M} - \mathcal{Q}  ) \left(\nabla \varphi_N \cdot \frac{\Dot{\mathbf{x}}}{c} \right)\frac{\Dot{\mathbf{x}}}{c} \qquad\qquad\qquad\quad\;\;\;\text{(MOG 1PN)} \\
            & - \alpha \nabla \varphi_N \left[ \left(2\mathcal{M} + \mathcal{Q}\right)\frac{\Dot{\mathbf{x}}^2}{2c^2}\right] \qquad\qquad\qquad\qquad\,\text{(MOG 1PN)} \\
            & - \alpha \nabla \varphi_N \left[
            \Big\{ 5 (\alpha + 1) \cancelto{\scriptstyle{0}}{(\mathcal{M} - \mathcal{Q})} + 4\mathcal{M} \Big\}\frac{\varphi_N}{c^2} \right]. \quad\text{(MOG 1PN)}
        \end{split}
        \end{align}
        We ordered the terms by powers in $1/c^{-2}$ and marked the physical interpretation of each line on the right. Moreover, recalling that the parameters $\mathcal{M}$ and $\mathcal{Q}$ are numerically equal to $1$, we stress the fact that $(\mathcal{M} - \mathcal{Q})$ is in reality $0$, but we included it to recall the book-keeping order of the expression.
        
        This derivation can be directly generalised to any modified theory of gravity, for instance, Yukawa gravity \citep{PhysRevD.109.044047}. We conclude this section with a few key observations.
        \begin{itemize}
            \item There is no MOG correction to the classical (Kepler) limit because the respective individual contributions of metric and fifth force cancel each other out. This is consistent with Eq.~\eqref{eq:MOG_weak_limit} in the limit $\mu_\phi r\gg 1$.
            \item Although the fifth force also counters the metric effect on the 1PN level to some extent, there are surviving terms. Therefore, although MOG is constructed to account for the missing mass problem via its weak-field limit Eq.~\eqref{eq:MOG_weak_limit}, its deviation from GR is also apparent in the strong-field regime.
            \item For $\alpha=0$, we recover the GR equations of motion up to 1PN order \citep{Gainutdinov_2020}, as it should be.
            \item We recall that Eq.~\eqref{eq:MOG_eom} constitutes an expression in isotropic coordinates (modulo the transformation to the Cartesian-like spatial basis).
        \end{itemize}
        
    \subsection{Formulation in terms of osculating Kepler orbits}\label{sec:pert_kepler}

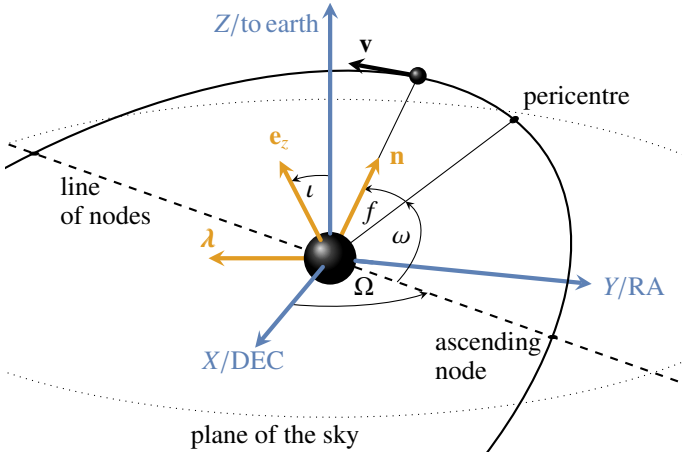
\begin{figure}

\tdplotsetmaincoords{70}{196}
\pgfmathsetmacro{\Om}{60}
\pgfmathsetmacro{\inc}{30}
\pgfmathsetmacro{\om}{75}
\pgfmathsetmacro{\f}{35}
\pgfmathsetmacro{\ompf}{110}
\pgfmathsetmacro{\Ompnz}{150}
\pgfmathsetmacro{\incmeps}{27}
\pgfmathsetmacro{\drawOm}{30}
\pgfmathsetmacro{\drawOmp}{175}
\pgfmathsetmacro{\drawom}{15}
\pgfmathsetmacro{\drawinc}{60}
\pgfmathsetmacro{\drawf}{125}
\pgfmathsetmacro{\aa}{.6}
\pgfmathsetmacro{\bb}{.5}
\pgfmathsetmacro{\cc}{.33166}
\pgfmathsetmacro{\dd}{.26833}
\pgfmathsetmacro{\rMBH}{.03}
\pgfmathsetmacro{\rstar}{.01}
\pgfmathsetmacro{\Xlength}{.31} \pgfmathsetmacro{\Xlabel}{.36}
\pgfmathsetmacro{\Ylength}{.31} \pgfmathsetmacro{\Ylabel}{.36}
\pgfmathsetmacro{\Zlength}{.31} \pgfmathsetmacro{\Zlabel}{.36}
\pgfmathsetmacro{\xlength}{.3} \pgfmathsetmacro{\xlabel}{.31}
\pgfmathsetmacro{\ylength}{.3} \pgfmathsetmacro{\ylabel}{.31}
\pgfmathsetmacro{\zlength}{.2} \pgfmathsetmacro{\zlabel}{.21}

\begin{tikzpicture}[scale=11.6,tdplot_main_coords]

\clip[tdplot_screen_coords] (-.37,-.22) rectangle (.40,.29);

\definecolor{mmblue}{rgb}{0.368417, 0.506779, 0.709798}
\definecolor{mmorange}{rgb}{0.880722, 0.611041, 0.142051}

\coordinate (O) at (0,0,0);
\coordinate (X) at (0,\Xlength,0); \coordinate (Xlabel) at (0,\Xlabel,0);
\coordinate (Y) at (-\Ylength,0,0); \coordinate (Ylabel) at (-\Ylabel,0,0);
\coordinate (Z) at (0,0,\Zlength);  \coordinate (Zlabel) at (0,0,\Zlabel);

\draw [-stealth,ultra thick,mmblue] (-\rMBH,0,0) -- (Y) node at (Ylabel){$Y/\mathrm{RA}$};

\tdplotsetrotatedcoords{\Om}{0}{0}
\begin{scope}[tdplot_rotated_coords]
\tdplotdrawarc[-stealth]{(O)}{.16}{\drawOm}{90}{anchor=south}{$\Omega$}
\draw[dotted] (O) circle (.53);
\node at (40:.6) {plane of the sky};
\draw[thick,dashed] (0,-.7,0) -- (0,.7,0);
\node at (99:-.32) [align=left] {line\\of nodes};
\draw[fill=black] (0,.365,0) circle (.005);
\node at (75:.372)[align=left]{ascending\\node};
\draw[fill=black] (0,-.485,0) circle (.005);
\end{scope}

\tdplotsetrotatedcoords{\Om}{\inc}{0}

\tdplotsetrotatedcoords{\Om}{\inc}{\om}
\tdplotdrawarc[tdplot_rotated_coords,-stealth]{(O)}{.11}{\drawom}{90}{anchor=east}{$\omega$}
\begin{scope}[tdplot_rotated_coords]
\tdplotdrawarc[-stealth]{(O)}{.11}{90}{\drawf}{anchor=north east}{$f$}
\draw[ultra thick,-stealth] (\drawf:.289) -- (143:.325) node [anchor=south west]{$\mathbf v$};

\draw[thick] (0,-\cc,0) ellipse (.5 and \aa);
\draw[fill=black] (0,\dd,0) circle (.005) node [anchor=south west] {$\mathrm{pericentre}$};
\draw (0,\rMBH,0) -- (0,\dd,0);
\end{scope}

\tdplotsetrotatedcoords{\Om}{\inc}{\ompf}
\begin{scope}[tdplot_rotated_coords]
\draw (0,\rMBH,0) -- (0,.287,0);
\draw [red,ultra thick,-stealth,mmorange,line cap=round] (0,\rMBH,0) -- (0,.16,0) node [anchor=west]{$\mathbf n$};
\draw [red,ultra thick,-stealth,mmorange,line cap=round] (-\rMBH,0,0) -- (-.16,0,0) node [anchor=south]{$\pmb{\lambda}$};
\end{scope}

\tdplottransformmainscreen{0}{0}{0}
\begin{scope}[tdplot_screen_coords]
\shade[ball color = black] (\tdplotresx,\tdplotresy) circle (\rMBH);
\shade[ball color = black] (.1,.208) circle (\rstar);
\end{scope}

\draw[tdplot_rotated_coords,ultra thick,-stealth,mmorange,line cap=round] (0,0,\rMBH) -- (0,0,.16) node [anchor=south]{$\mathbf e_z$};

\tdplotsetthetaplanecoords{0}
\tdplotdrawarc[tdplot_rotated_coords,-stealth]{(0,0,0)}{.1}{0}{\incmeps}{anchor=north}{$\iota$}

\tdplotsetrotatedcoords{0}{0}{0}
\begin{scope}[tdplot_rotated_coords,ultra thick,mmblue]
\draw [-stealth,line cap=round](0,\rMBH,0) -- (X) node at (Xlabel){$X/\mathrm{DEC}$};
\draw [line cap=round](-\rMBH,0,0) -- (-.05,0,0);
\draw [-stealth,line cap=round](0,0,\rMBH) -- (Z) node [anchor=north east]{$Z/\mathrm{to\,earth}$};
\end{scope}

\end{tikzpicture}

\caption{Definition of the Euler angles $(\Omega,\iota,\omega)$ giving the orbital orientation and of the true anomaly $f$ giving  the current position within the orbit. The static fundamental frame $(X,Y,Z)$ (blue) is adapted to the observables, while the co-rotated Gaussian frame $(\mathbf{n}, \pmb{\lambda}, \mathbf{e}_z)$ (orange) is adapted to the symmetry of the (perturbed) two-body problem (reproduced from~\citet{Heissel+22}).}
\label{F: orientation angles}
\end{figure}

        Motivated by the final point of the previous section, we prepared Eq.~\eqref{eq:MOG_eom} for an investigation into the effect of MOG on an orbit that probes the PN regime of a central mass $M$. To do this, we viewed our equations of motion as a perturbed Kepler problem,
        \begin{equation}\label{eq:pertubed_kepler}
            \Ddot{\mathbf{x}} = -\frac{G_N M}{\rho^2}\mathbf{n} + \mathbf{a}_{\text{p}},
        \end{equation}
        where the perturbative acceleration $\mathbf a_p$ abbreviates all terms on the right-hand side of Eq.~\eqref{eq:MOG_eom} but the first one,  $\rho=|\mathbf x|$ and $\mathbf n=\mathbf x/\rho$. We still kept track of the terms proportional to $(\mathcal M - \mathcal Q)$ for later utility. We projected $\mathbf a_p$ onto the Gaussian frame \citep{PoissonWill2014, Merritt2013},
        \begin{equation}
            \mathbf{a}_{\text{p}} = \mathcal{R} \mathbf{n} + \mathcal{S} \pmb{\lambda} + \mathcal{W} \mathbf{e}_z\, ,
        \end{equation}
        where $\mathbf{e}_z$ is the unit vector aligned with the orbital angular momentum, and $\pmb{\lambda}$ completes the right-handed orthonormal basis (Fig.~\ref{F: orientation angles}).
        The components can be further expressed as
        \begin{subequations}\label{eq:RSW_comps}
        \begin{align}
            \mathcal{R} &= \;\,\mathcal{R}_{\text{GR\,1PN}} + \;\,\mathcal{R}_{\text{MOG\,1PN}} \, ,\label{eq:R_component}\\
            \mathcal{S} &= \;\,\mathcal{S}_{\text{GR\,1PN}} + \;\,\mathcal{S}_{\text{MOG\,1PN}} \, , \\
            \mathcal{W} &= \mathcal{W}_{\text{GR\,1PN}} + \mathcal{W}_{\text{MOG\,1PN}} \, ,
        \end{align}
        \end{subequations}
        where the GR 1PN components read \citep{Merritt2013}
        \begin{subequations}\label{eq:RSW_GR_comps}
        \begin{align}
            &\mathcal{R}_{\mathrm{GR\,1PN}} = \frac{G_N M}{c^2 \rho^2} \left[ \frac{4 G_N M}{\rho} - \Dot{\mathbf{x}}^2 + 4(\mathbf{n}\cdot \Dot{\mathbf{x}})^2 \right] \, ,\\
            &\mathcal{S}_{\mathrm{GR\,1PN}} = \frac{4 G_N M}{c^2 \rho^2} (\mathbf{n}\cdot \Dot{\mathbf{x}}) (\pmb{\lambda}\cdot \Dot{\mathbf{x}})\, ,\\
    &\mathcal{W}_{\mathrm{GR\,1PN}}= 0\, ,
        \end{align}
        \end{subequations}
    with
    \begin{subequations}
        \label{eqn_expl}
        \begin{align}
            &\rho = \frac{p}{1 + e \cos{f}} =  \frac{a ( 1-e^2)}{1 + e \cos{f}}, \\
           & \dot{\mathbf x}^2  = \frac{G_N M}{p} (1 + e^2 + 2 e \cos{f}) , \\
            &(\mathbf{n}\cdot \Dot{\mathbf{x}}) = \sqrt{\frac{G_N M}{p}} e \sin{f} , \\
            &(\pmb{\lambda}\cdot \Dot{\mathbf{x}}) = \sqrt{\frac{G_N M}{p}} (1 + e \cos{f}).
        \end{align}
        \end{subequations}
     For the MOG components, we derived        \begin{subequations}\label{eq:RSW_MOG_comps}
        \begin{align}
            \begin{split}
                &\mathcal{R}_{\text{MOG}} = -\alpha \frac{G_N M}{\rho^2} \cancelto{\scriptstyle{0}}{(\mathcal{M} - \mathcal{Q})} \\
                &\phantom{\mathcal{R}_{\text{MOG}}}\quad
                + \alpha \frac{G_N M}{\rho^2}
                \left\{(4\mathcal{M} - \mathcal{Q})\frac{(\mathbf{n}\cdot \Dot{\mathbf{x}})^2}{c^2} \right. - (2\mathcal{M} + \mathcal{Q}) \frac{\Dot{\mathbf x}^2}{2c^2} \\
&\phantom{\mathcal{R}_{\text{MOG}}}\quad \left.+ \frac{G_N M}{c^2\rho}
                \left[  5(\alpha+1)\cancelto{\scriptstyle{0}}{(\mathcal{M} - \mathcal{Q})}+4 \mathcal{M} \right] \right\}\, ,
            \end{split}
            \\
            &\mathcal{S}_{\text{MOG}} = \alpha (4\mathcal{M} - \mathcal{Q}) \frac{G_N M}{c^2 \rho^2} (\mathbf{n}\cdot \Dot{\mathbf{x}}) (\pmb{\lambda}\cdot \Dot{\mathbf{x}}) \, ,
            \\
    &\mathcal{W}_{\text{MOG\,1PN}} = 0\, .
            \label{eq:S_MOG_component}
        \end{align}
        \end{subequations}
         Following \cite{PoissonWill2014}, we used the formalism of osculating orbits, in which the perturbed Kepler problem (given by Eq.~\eqref{eq:pertubed_kepler}) is reformulated as a set of evolution equations for the six orbital elements $(p,e,\iota,\Omega,\omega, \text{and }f)$ correspondent to semi-latus rectum, eccentricity, inclination, argument of the ascending node, argument of pericentre, and true anomaly, respectively (see Fig.~\ref{F: orientation angles}). The full equations are explicitly reported in Eqs.~(3.64) of~\cite{PoissonWill2014}, while they are here written for the special case $\mathcal W = 0$,   \begin{subequations}\label{eq:osc_eqs}
        \begin{align}
            \dot p &= \sqrt{\frac{p^3}{G_N M}}\frac{2}{1+e \cos f}\mathcal S \, , \\
            \dot e &= \sqrt{\frac{p}{G_N M}}\left[\sin f\,\mathcal R+\frac{2\cos f+e(1+\cos^2f)}{1+e\cos f}\mathcal S\right] \, ,\\
            \dot\omega &= \frac{1}{e}\sqrt{\frac{p}{G_N M}}\left[-\cos f\,\mathcal R+\frac{2+e\cos f}{1+e\cos f}\sin f\,\mathcal S\right] \, ,
        \end{align}
        \end{subequations}
        while $\dot\iota = 0$ and $\dot\Omega = 0$. For our problem, the remaining Gaussian frame components $\mathcal R$ and $\mathcal S$ are given by Eqs.~\eqref{eq:RSW_comps}, \eqref{eq:RSW_GR_comps} and \eqref{eq:RSW_MOG_comps}. Finally, we obtained closed-form expressions by using the relations \eqref{eqn_expl}.

    At any point in time, the transformation between the orbital elements, position, and velocity is given by Eqs.~\eqref{eq:pos_and_vel}. Some qualitative features of the MOG orbital mechanics to 1PN order can be seen directly from Eqs.~\eqref{eq:osc_eqs}:
    \begin{itemize}
        \item  Because of the vanishing out-of-plane component of the perturbative acceleration $\mathcal W$, the elements $\iota$ and $\Omega$ are constants of motion, such that MOG does not alter the orbital plane.
        \item  Because of the non-vanishing in-plane components of the perturbative acceleration $\mathcal R_\mathrm{MOG}\,$, $\mathcal S_\mathrm{MOG}$ (Eqs.~\eqref{eq:RSW_MOG_comps}), MOG contributes to the evolution of $p$, $e$, and $\omega$. The latter implies in particular that it contributes to the rotation of the orbit within its plane.
    \end{itemize}

    \subsection{Secular evolution}\label{sec:secular_evolution}

        Next, we investigated whether the evolutions of $p$, $e$, and $\omega$ exhibited secular components, that is, whether there are non-vanishing changes in these elements per revolution, for instance, $\Delta\omega=\omega(f+2\pi)-\omega(f) $, and hence, accumulating effects over several orbits. To do this, we assumed the trajectories to be close to Keplerian, which is valid on the distance scales in which we are interested. Then, we considered the first-order approximation of the evolution Eqs.~\eqref{eq:osc_eqs}       \begin{subequations}\label{eq:osc_eqs_first_order},
        \begin{align}
            \frac{dp}{df}&\simeq
            \frac{p^3}{G_NM}\frac{2}{(1+e \cos f)^3}\mathcal S
            \\
            \frac{de}{df}&\simeq
            \frac{p^2}{G_NM}\Bigg[\frac{\sin f}{(1+e \cos f)^2}\mathcal R + \frac{2\cos f + e(1+\cos^2 f)}{(1+e \cos f)^3}\mathcal S\Bigg]
            \\
            \frac{d\omega}{d f}&\simeq \frac{p^2}{G_N M}\frac{1}{e}
            \Bigg[ -\frac{\cos f}{(1+e \cos f)^2}\mathcal{R}
            + \frac{2 + e \cos f}{\big(1+e \cos f\big)^3} \sin f \mathcal{S}\Bigg] \label{eq:domega_df},
        \end{align}
        \end{subequations}
        in which the orbital elements on the left-hand side are viewed as functions of $f$, all orbital elements on the right-hand side except $f$ are assumed constant, and $\dot f$ is approximated by its non-perturbed Keplerian expression \citep[p.~160]{PoissonWill2014}.
        
        Plugging in $\mathcal{R}$ and $\mathcal{S}$ from Eqs.~\eqref{eq:RSW_comps}--\eqref{eq:RSW_MOG_comps}, we then integrated Eqs.~\eqref{eq:osc_eqs_first_order} analytically over one period. More precisely, the secular change of an orbital element $\mu^a$ over a complete orbit is given by
        \begin{equation}
            \label{eq:Delta_orb}
            \Delta \mu^a =\int_{0}^{P} \frac{d \mu^a}{dt}\, dt =\int_{0}^{2\pi} \frac{d \mu^a}{df} \,df\, ,  
         \end{equation}
        where $P$ is the orbital period. Thus, we found $\Delta p=0$, $\Delta e=0$, and
        \begin{equation}\label{eq:delta_om_2pi}
            \Delta \omega = \frac{6 \pi G_N M}{p c^2} \left(1 + \frac{5}{6}\alpha\right).
        \end{equation}
        
        Moreover, for an orbital element $\mu^a_i$, denoting a combination of effects $i$ (as $i\in\{\mathrm{GR},\,\mathcal{ST},\,\mathcal{FF}\, , \ldots\}$, see Table~\ref{tab:MOG_contributions}), we defined $\delta \mu^a_i$ as
        \begin{equation}\label{eq:delta_orb}
           \delta \mu^a_i (f) \equiv \Delta \mu^a_i (f) - \Delta \mu^a_{\text{GR}}(f).
        \end{equation}
        This is useful in our discussion of Fig.~\ref{fig:effects_on_orbital_elements} below.
        
        This shows that
        \begin{itemize}
            \item there is no secular evolution of $p$ and $e$, such that these elements return to their original values after each revolution,
            \item the only element other than $f$ that exhibits a secular evolution is $\omega$, and hence, there is a precession of the orbit within its plane. For $\alpha=0$, we recover the GR expression, famous ever since its first derivation by \cite{Einstein:1915bz}. For $\alpha>0$, MOG amplifies this precession by the above factor. 

            \item We note, however, that MOG and GR can both very well yield the same $\Delta\omega$ since the gain in precession by an $\alpha>0$ can be counterbalanced by a lower central mass. Through correlations like this, that knowledge of the secular behavior alone is insufficient to distinguish between different theories of gravity and dark matter. To do this, an understanding of the distinct non-secular signatures of each model is key (see Sect.~\ref{sec:MOG_impacts} and~\citet{Heissel+22}.
            
            \item Our result (Eq~\eqref{eq:delta_om_2pi}) confirms the result of \citet{Della_Monica_2021, 10.1093/mnras/stad579}, who performed an alternative derivation. In Sect.~\ref{appendix:ffandem} we also settle a disagreement between these authors and \cite{10.1093/mnras/stac2113}.
        \end{itemize}

    \subsection{Observational effects: Transverse Doppler shift and gravitational redshift}\label{sec:redshift_formula}

        When the interest is not only in the dynamics of test particles around a MOG black hole as such, but furthermore, in how this dynamics would be observed from a distance, MOG effects on the light by which we observe have to be considered as well. We restricted our considerations here to two such effects, namely the transverse Doppler shift and the gravitational redshift, since these are currently the only relativistic effects of relevance to the observed light of star S2 in the Galactic centre. Moreover, we consider this latter system as the example to investigate MOG features in the non-secular evolution in Sect.~\ref{sec:MOG_impacts}.
        
        First, we note that as a special relativistic effect, the transverse Doppler effect is unaltered by MOG. Furthermore, since photons are massless, there is no fifth-force contribution to the gravitational redshift. The MOG BH metric~(see Eqs.~\eqref{eq:MOG_line_element}-- \eqref{eq:psi_MOG_1}), however, does yield modifications to the well-known GR redshift. Labelling the event at which a photon is emitted at the star by $P$ and the event at which this photon is detected on Earth by $Q$, we conclude according to \cite{Misner:1973prb} that for the gravitational redshift $z$, we have
        \begin{equation}
            1+z =
            {\frac{\nu_P}{\nu_Q}} = \sqrt{\frac{g_{00}|_{ Q}}{g_{00}|_{ P}}} =
            \sqrt{\frac{\Psi(|\mathbf{x}_{ Q}|)}{\Psi(|\mathbf{x}_{ P}|)}} =
            \sqrt{\frac{\Psi({R})}{\Psi({\rho})}},
        \end{equation}
        where $\nu_P$, $\nu_Q$ are the emitted and observed frequencies, respectively, $R$ is the distance from Earth to the Galactic centre, and $\Psi$ is given by Eq.~\eqref{eq:psi_MOG_1} transformed into isotropic coordinates (see Eq.~\eqref{eq:Psi_f2}). On the other hand, the radial (line-of-sight) velocity, namely RV, is inferred from redshift via
        $$1+z=\sqrt{\frac{1+\Delta RV/c}{1-\Delta RV/c}}\approx 1+ \frac{\Delta RV}{c}$$
        for $|\Delta RV|\ll c\,$, by which we obtain
        \begin{align}\label{eq:redshift}
            \frac{\Delta RV}{c} \approx \sqrt{\frac{1 - 2 (\alpha + 1)\frac{G_N M}{{R_0} c^2}}{1 - 2 (\alpha + 1)\frac{G_N M}{{\rho} c^2}}} - 1
        \end{align}
        due to the gravitational redshift, where we kept $\Psi$ up to 1PN order. Therefore, in order to obtain the radial velocity that the star had at emission, we need to subtract the $\Delta RV$ (given by Eq.~\eqref{eq:redshift}) from the $RV_Q$ measured on Earth, obtaining $RV_P = RV_Q - \Delta RV$. For $\alpha=0$, Eq.~\eqref{eq:redshift} reduces to the familiar result of GR.

\section{Effects of MOG on stellar orbits in the Galactic centre}\label{sec:MOG_impacts}

    \subsection{Model setup: Star S2 in orbit around Sagittarius A*}

    For the system in the example of which we conducted our numerical investigations, we chose star S2, which is in orbit around the massive black hole Sgr~A* in the Galactic centre \citep{Ghez+08,GillessenEtAl2009, GRAVITY+22_mass_distribution}. With a pericentre distance of about $1400$ Schwarzschild radii, it probes the post-Newtonian regime of Sgr~A*. Coupled with a high eccentricity of about $0.88$, its orbit serves as an excellent laboratory for testing relativity \citep{GRAVITY+18_redshift, GRAVITY+20_Schwarzschild_prec, Do+19, SaidaEtAl2019}, dark matter \citep{Heissel+22,10.1038/s41598-022-18946-7,PhysRevD.108.L101303,GRAVITY+22_mass_distribution}, and alternative theories of gravity \citep{Borka_2013,PhysRevD.104.L101502,universe8020137,10.1093/mnras/stac3648,GRAVITY:2025ahf}.
    
    The hypothetical scenario in which Sgr~A* is assumed to be a MOG black hole was investigated by \citep{Della_Monica_2021}, who obtained the upper bound $\alpha<0.662$ with a confidence of $99.7\%$. The measurement was based on fitting the fully relativistic equations of motion (Eq.~\eqref{eq:MOG_compact_eom}) to publicly available data of astrometric and spectroscopic S2 observations. We instead investigated which distinct signature a MOG black hole inscribes on the S2 orbit as compared to a Schwarzschild black hole, and we compared the astrometric and spectroscopic observables, as well as the evolution of the orbital elements. We hence based our model on the post-Newtonian osculating-orbit framework discussed in Sect.~\ref{sec:pert_kepler}. This choice also facilitates comparisons to dark matter scenarios (see Sect.~\ref{sec:MOG_vs_DM}) and prepares MOG for potential future Solar System tests, in which Newtonian dominate relativistic perturbations, such that a fully relativistic model is practically impossible to set up.

    In the following, model orbits are based on the integration of the osculating Eqs.~\eqref{eq:osc_eqs} with $\mathcal{R,S}$ given by Eqs.~\eqref{eq:RSW_comps}--\eqref{eq:RSW_MOG_comps}. For the initial point ($t_0$), we chose the (at the time of writing) last apocentre passage of the star in $2010$. For the other initial orbital elements, the black hole mass and the distance to the Galactic centre, we chose the same values as in~\citet[Table~B.1]{GRAVITY+22_mass_distribution}. In summary, our model parameters common to all orbits were
    \begin{subequations}\label{eq:model_params}
    \begin{align}
        p_0 &= 0.00109214\,\mathrm{pc}\, , &
        e_0 &= 0.88441 \, ,&
        \iota_0 &= 134.70^\circ \, ,\\
        \omega_0 &= 66.25^\circ \, , &
        \Omega_0 &= 228.19^\circ \, ,&
        f_0 &= \pi \, ,\\
        M &= 4.297\times10^6\,M_\odot\, ,&
        R &= 8277\,\mathrm{pc}\, ,
    \end{align}
    \end{subequations}
    where the orbital elements correspond to the initial time $t_0$. The sole purpose of the distance to the Galactic centre $R$ is to translate between physical distances and angular distances on sky. In one investigation, which we point out explicitly, we chose different values for the black hole mass. The parameters that we varied to investigate the theory were
    \begin{align}
        \alpha\geq0,\quad
        \mathcal M\in\{0,1\}\,\,\,\text{ and }\,\,\,
        \mathcal Q\in\{0,1\},
    \end{align}
    where the latter two were set to $1$, unless stated otherwise, and were occasionally set to $0$ individually in order to investigate the fifth force from the MOG space-time contribution in isolation.
        
    Since we did not fit observations, but focused on features of MOG, we did not model systematic effects such as the motion of the Solar System or the offset and drift of instrument reference frames (as in e.g. \cite{GRAVITY+20_Schwarzschild_prec} and~\cite{ReidBrunthaler20}). Moreover, in comparison plots, where we subtract a quantity of a GR orbit from the corresponding quantity of a MOG orbit, these contributions would cancel out in any case because they affect the two orbits identically. To investigate the effects on the observables $\mathrm{(RA,DEC,and~ RV)}$, which are the right ascension, declination, and radial velocity, we transformed the orbital elements in position and velocity using Eqs.~\eqref{eq:pos_and_vel}, such that $\mathrm{(RA,DEC,RV)} = (r^Y,r^X,-v^Z)$ (Fig.\mm{~\ref{F: orientation angles}}).    
    
    \subsection{Comparing MOG orbits to GR orbits}
\begin{figure}
            \centering
            \includegraphics[width=0.99\linewidth]{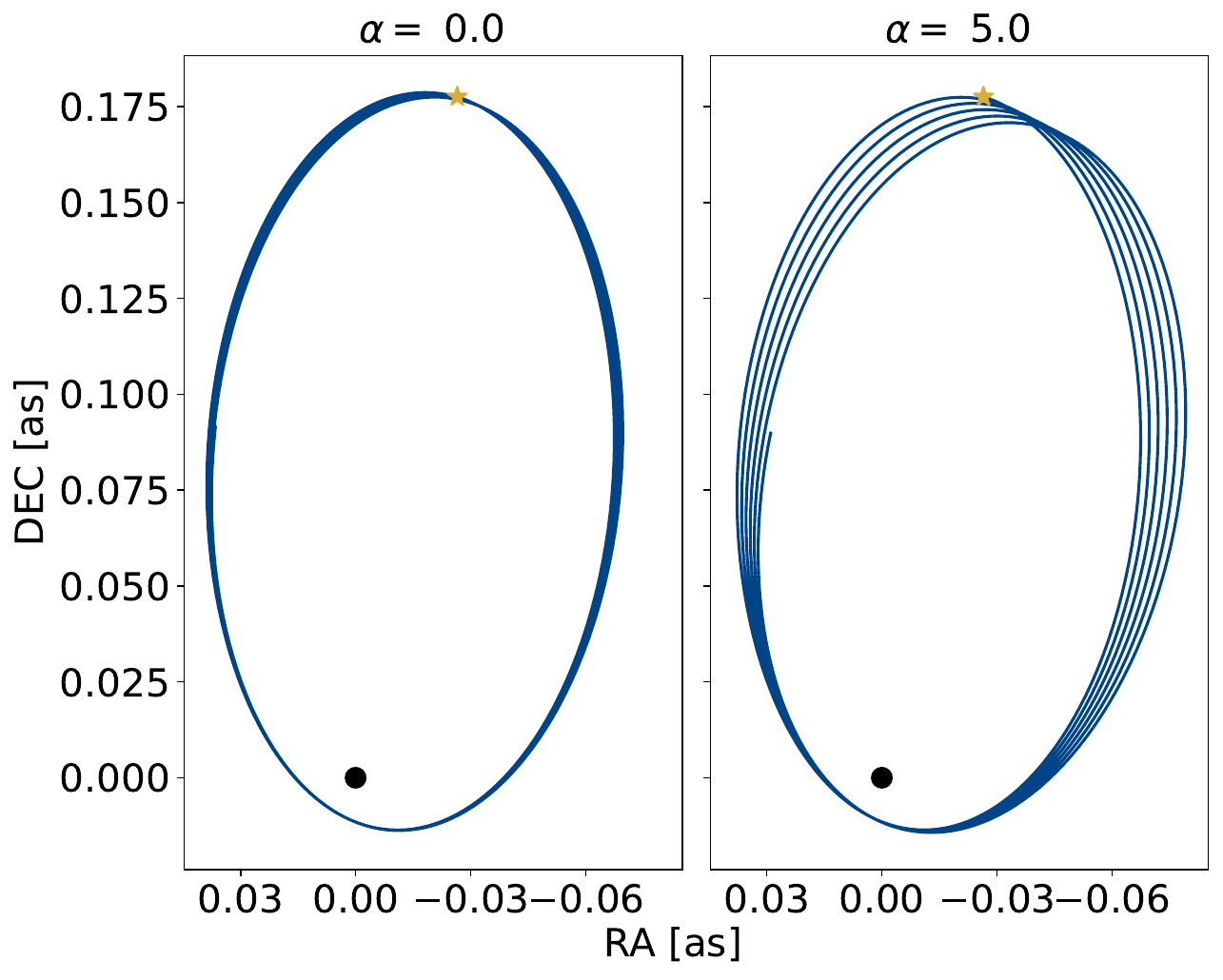}
            \caption{Oribt of star S2 (yellow star) around the BH (black dot) in the centre of our Galaxy for GR (left panel) and an extreme MOG case (right panel).}
            \label{fig:extreme_MOG}
        \end{figure}

        In Sect.~\ref{sec:secular_evolution} we determined that for the same central mass $M$, MOG enhances the secular pericentre advance of GR (Schwarzschild precession) by a factor $1+5\alpha/6$ (Eq.~\eqref{eq:delta_om_2pi}). Figure~\ref{fig:extreme_MOG} demonstrates how this alters the appearance of the observed orbit on sky for the example value $\alpha = 5$, yielding an almost sixfold Schwarzschild precession.
        Since publicly available observational data readily constrain $\alpha< \alpha_{\ast}:=\, 0.662$ \citep{Della_Monica_2021}, this is an extreme choice for the sole demonstrative purpose in which the effect is visible by naked eye.
        
\begin{figure*}
            \centering
            \includegraphics[width=0.99\linewidth]{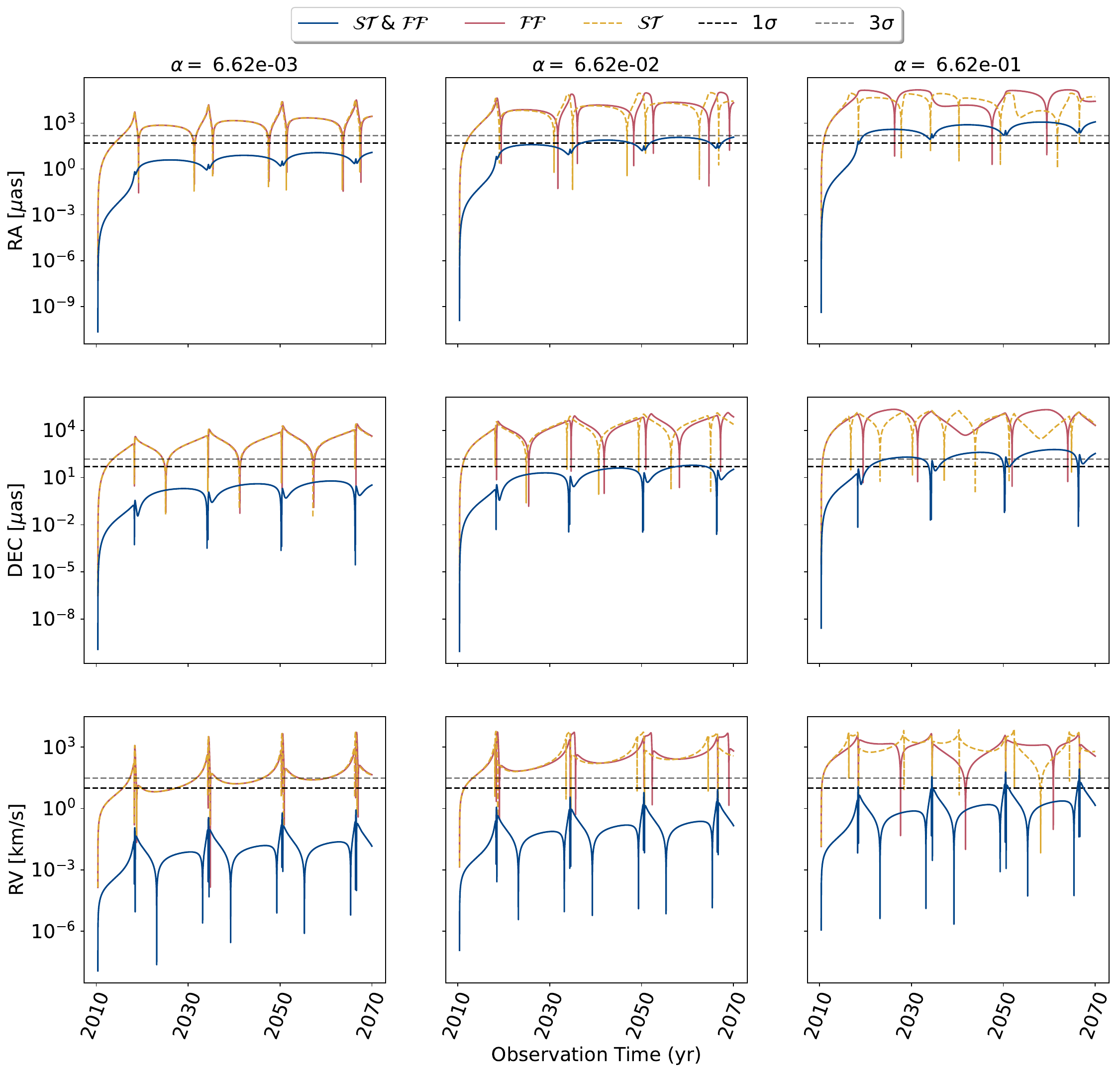}
            \caption{Evolution of the observables' (RA, DEC, and RV in the first, middle, and bottom row, respectively) absolute difference between GR and i) full MOG (solid blue lines), ii) fifth force (solid red lines), and iii) space-time (dashed yellow lines) contribution for different values of $\alpha\,$. We fixed $\alpha$ as $\alpha=6.62\cdot 10^{-3}$ in the left column, $\alpha=6.62\cdot 10^{-2}$ in the middle one and $\alpha=6.62\cdot 10^{-1}$ in the right column. An explanation of the labels can be found in Table \ref{tab:MOG_contributions}. The horizontal dashed lines correspond to the $1\sigma$ (black) and $3\sigma$ (grey) accuracy of the instruments. For the spatial observables (RA and DEC), these are $ 50 \ \mathrm{\mu as}$ and $150 \ \mathrm{\mu as}$, respectively, and for the radial velocity (RV), they were fixed to $10 \mathrm{\ km/s}$ and $30 \mathrm{\ km/s}$.
            }
            \label{fig:observables}
        \end{figure*}

        To place into perspective what is actually observable, we note that the current $1\sigma$ measurement uncertainties for S2 observations are about $50\,\mathrm{\mu as}$ in astrometry using the GRAVITY instrument at the Very Large Telescope Interferometer \citep{GRAVITY+17_first_light} and $10\,\mathrm{km/s}$ in radial velocity using the Spectrograph for Integral Field Observations in the Near Infrared \citep{Eisenhauer_2005} and the Enhanced Resolution Imager and Spectrograph \citep{refId0} at the Very Large Telescope. Therefore, in order to probe the detection threshold for the deviation of a MOG orbit from a GR orbit for realistic $\alpha$ values, we present Fig.~\ref{fig:observables}.
        We subtracted the observables of the GR orbit from the observables of the MOG orbit on a logarithmic scale as functions of time for three values of $\alpha$, the highest of which is the bound $\alpha_{\ast}$ found by \citet{Della_Monica_2021}.
        By our choice of initial parameters (Eq.~\eqref{eq:model_params}), the orbits coincide at the $2010$ apocentre passage and then deviate with a certain signature within each $16$-year period and an overall accumulation of the deviation per revolution according to Eq.~\eqref{eq:delta_om_2pi}.
        
        The precise form of the non-secular deviation has to be interpreted with some caution. In Fig.~\ref{fig:observables}, which shows the observables in the time domain, the sharp features around pericentre are in part due to the fact that the orbital period changes with $\alpha$. This ought to be automatically corrected in the true anomaly domain. A certain $f$ value might correspond to different times $t$ for orbits in different gravity theories. Furthermore, due to the high eccentricity we considered (see Eq.~\eqref{eq:model_params}), the part of the orbit around the pericentre is traversed in a rather short window in the time domain, which is stretched out slightly further in the true anomaly domain (see the grey area in Fig.~\ref{fig:DMvsMOG}).

        Overall, we observe that for $\alpha$ equal to the current bound (right column of Fig.~\ref{fig:observables}), the MOG deviation from GR (solid blue lines) surpasses the $1\sigma$ instrument threshold (dashed horizontal line) after a single revolution ($\approx$ 16 years), and the $3\sigma$ limit (horizontal black line) shortly after. For lower $\alpha$ values (left and middle columns of Fig.~\ref{fig:observables}), the effect has to accumulate for longer in order to exceed the thresholds. However, an improvement of the \citep{Della_Monica_2021} limit by a factor $1/10$, namely $\alpha=6.62\cdot 10^{-2}$, represented in the middle column of Fig.~\ref{fig:observables}, seems achievable within realistic timescales.
        
        Furthermore, the $\mathrm{RA}$ deviation has a positive offset with respect to the $\mathrm{DEC}$ deviation because of the spatial orientation of the orbit. Finally, we also note that the $\mathrm{RV}$ deviation needs to accumulate longest in order to breach the thresholds. However, we recall that the deviation plotted here is solely due to the effects of MOG on the orbital motion itself. In addition to this, spectral observations are subject to the gravitational redshift, which is altered by MOG in comparison to GR as well (Sect.~\ref{sec:redshift_formula}). We discuss the MOG redshift contribution to $\mathrm{RV}$ in isolation in Sect.~\ref{sec:redshift}.

\begin{figure*}
            \centering
            \includegraphics[width=0.99\linewidth]{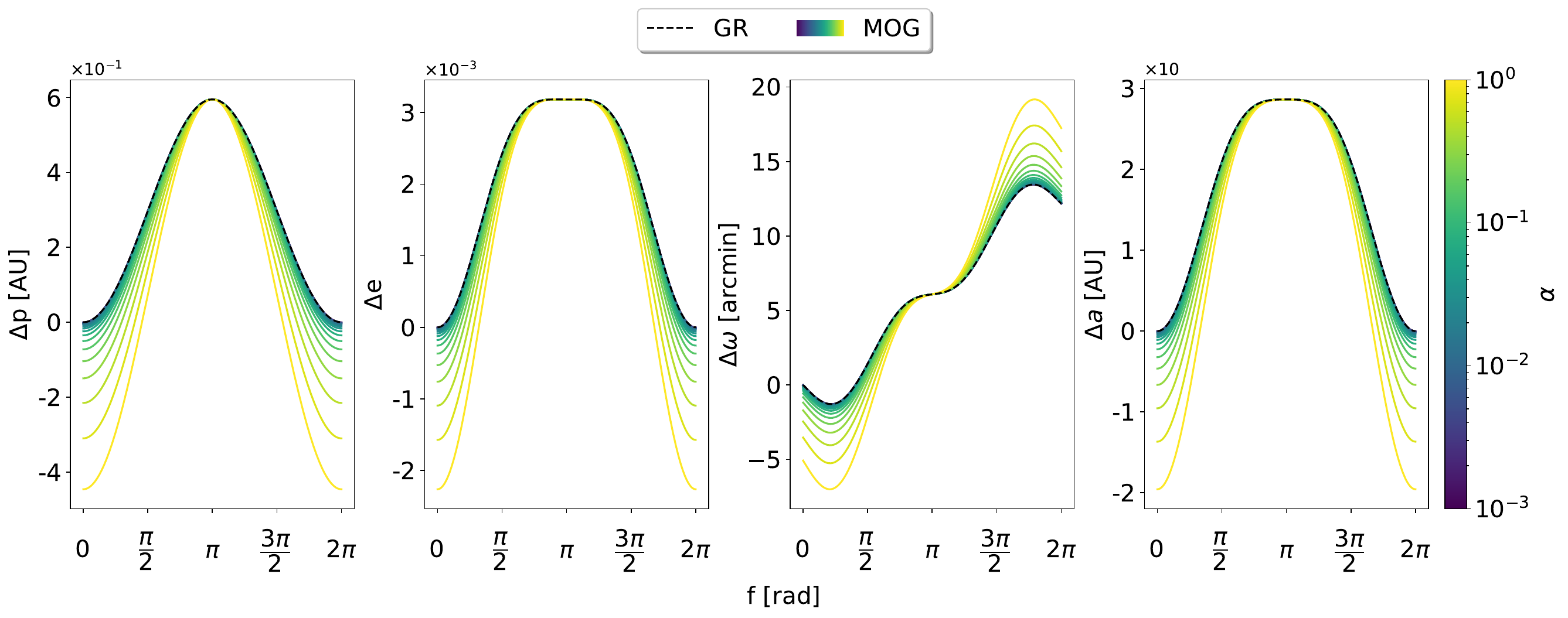}
            \caption{Orbital element evolution with true anomaly. Here, $\Delta p$ stands for the offset of $p$ from its GR value at $f=0$, and analogously for the other elements. The dashed black line represents GR, and the solid coloured lines represent a MOG case with a particular $\alpha$ value in the range ($0.001,1$). For each orbital element $\mu^a$, the definition of $\Delta$$\mu^a$ is given in Eq.~\eqref{eq:Delta_orb}.}
            \label{fig:orbital_elements}
        \end{figure*}

        We finally focused on the representation of the orbits in terms of orbital elements $(p, e, \omega, a)\,$ and show their change over one revolution in Fig.~\ref{fig:orbital_elements}.
        Firstly, we note that the secular evolution agrees with what we derived in Sect.~\ref{sec:secular_evolution}, such that $\Delta p=\Delta e=0$, and $\omega$ changes from $f=0$ to $2\pi$ according to Eq.~\eqref{eq:delta_om_2pi}. Secondly, the coincidence of all orbits at $f=\pi$ is again due to our arbitrary choice of initial parameters (Eq.~\eqref{eq:model_params}). The absolute deviation between the MOG and GR orbits thus must not be interpreted as an indicator of where (in terms of true anomaly) MOG impacts the orbit. Instead, the slope of these curves marks where deviations from Keplerian motion are driven by both GR and MOG.
    
    \subsection{Fifth force and metric effects in isolation}\label{sec:FFvsST}
    
\begin{figure*}
            \centering
            \includegraphics[width=0.99\linewidth]{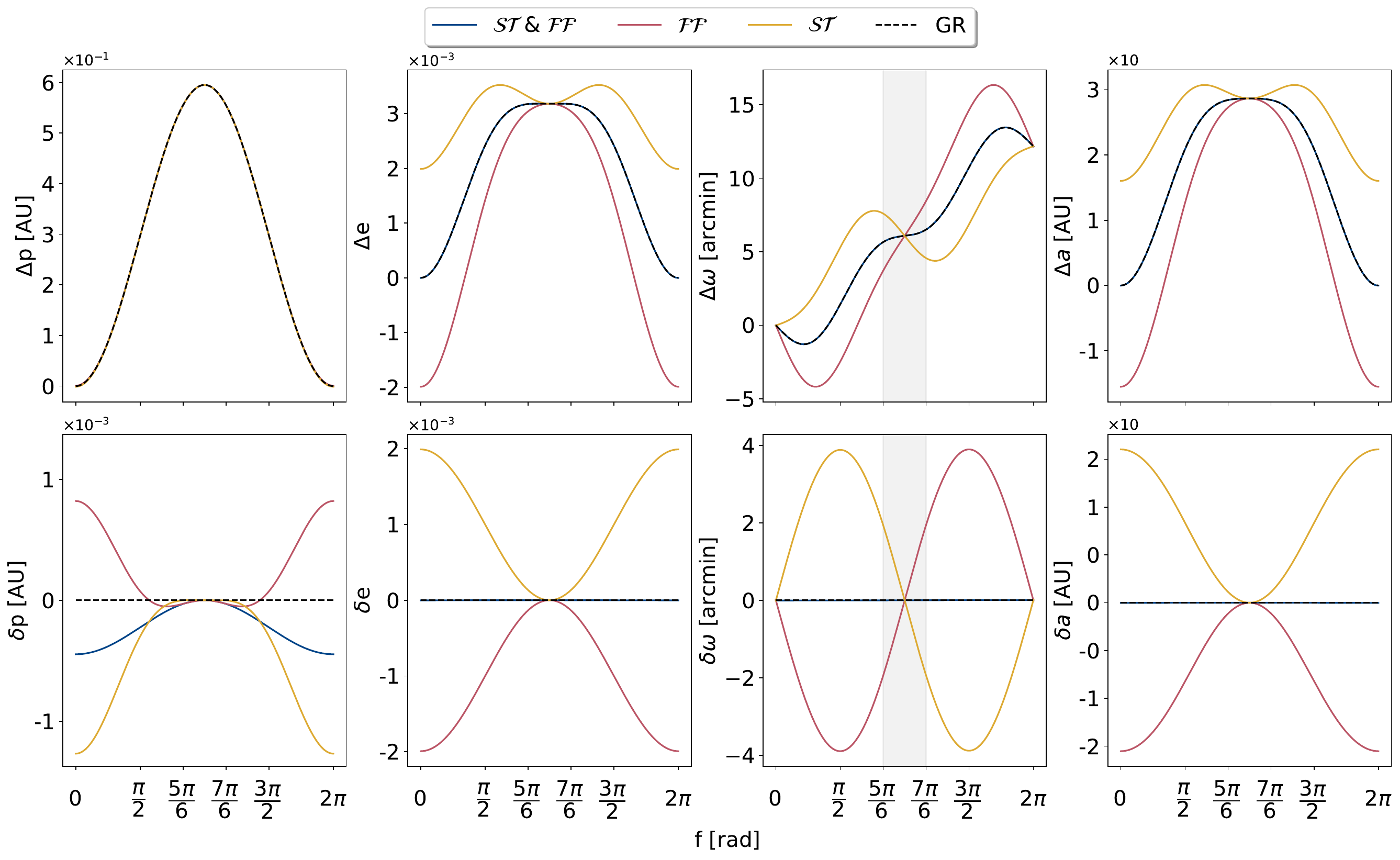}
            \caption{Effects on the evolution of orbital elements (top panel) and the deviation from GR (bottom panel). Each case is integrated independently. The cases are i) full MOG (solid blue line), ii) fifth-force contribution alone (solid red line), iii) space-time contribution alone (solid yellow line), and iv) GR (dashed black line). Cases ii) and iii) are not physical in MOG, but are used for diagnostic purposes, as explained in Sec.~\ref{sec:FFvsST}. An explanation of the labels can be found in Table \ref{tab:MOG_contributions}. For each orbital element $\mu^a$, the definitions of $\Delta$ $\mu^a$ and $\delta \mu^a$ are given in Eq.~\eqref{eq:Delta_orb} and~\eqref{eq:delta_orb}, respectively.
            }
            \label{fig:effects_on_orbital_elements}
        \end{figure*}

        As was already apparent from the fully relativistic equations of motion (Eqs.~\eqref{eq:MOG_compact_eom}), MOG accelerates massive test particles via two components: through its distinct metric from GR, and due to the fifth force, by which the motion is non-geodesic. In Sect.~\ref{sec:PN dynamics} we kept track of these separate contributions throughout our derivations of the PN equations of motion (Eq.~\eqref{eq:MOG_eom}) and of the perturbative acceleration components $\mathcal{R,S,\text{and } W}$ (Eq.~\eqref{eq:RSW_comps}) via the book-keeping parameters $\mathcal{M,Q}$. We used these then to investigate the individual contributions to the dynamics in isolation by setting them to zero individually. We stress that this is merely a diagnostic tool and does not reflect a physical scenario in the MOG theory, since any massive object carries a fifth-force charge. Therefore, the only physical scenario is the full MOG, or equivalently, $\mathcal{ST}$ \& $\mathcal{FF}$\: ($\mathcal{(M,Q)}=(1,1)$), and GR ($\mathcal{(M,Q)}=(0,0)$). However, due to the relation to Reissner-Nordström, shown in Sec.~\ref{sec:reiss_nord}, the $\mathcal{ST}$-only scenario around a MOG BH can be considered equivalent to a massive but electrically neutral particle following a geodesic around an electrically charged BH in GR.
        \begin{table}[htp]
            \centering
            \caption{MOG contributions and on-off parameters.}
            \begin{tabular}{|c|c|c|c|}
                \hline
                Contributions   & $\mathcal{M}$ & $\mathcal{Q}$ & Extended mass \\
                \hline
                GR              & 0             & 0             & -       \\
                $\mathcal{ST}$  & 1             & 0             & -       \\
                $\mathcal{FF}$  & 0             & 1             & -       \\
                $\mathcal{ST}$
                \&  
                $\mathcal{FF}$  & 1             & 1             & -       \\
                DM              & 0             & 0             & Plummer profile  \\
                \hline
            \end{tabular}
            \label{tab:MOG_contributions}
        \end{table}
        In order to select an $\alpha$ value for this investigation, we recall that for $\mathcal{(M,Q)}=(1,0)$ or $(0,1)$, the weak-field acceleration of MOG does not cancel, such that the Keplerian two-body acceleration is modified by a factor $1\pm\alpha$ (first lines in Eqs.~\eqref{eq:MOG_eom} and~\eqref{eq:RSW_MOG_comps}). Being of 0PN order, this term then completely dominates the perturbative accelerations, such that we had to choose a very small $\alpha$ to yield curves on reasonable scales. We chose $\alpha=10^{-3}$. For the same reason, we waited for the isolated effects to appear equal and opposite in our plot, up to small 1PN differences not visible by naked eye on these scales for such a small $\alpha$. Consequently, the difference between the curves for GR (dashed black) and the full MOG (solid blue) theory are also expected to be insignificant and overlap by naked eye. All of the above can indeed be observed in Fig.~\ref{fig:effects_on_orbital_elements}.
        
        The semi-latus rectum $p$ (first column of the figure) does not strongly deviate from GR even for single contributions. The maximum effect is about $10^{-3}$ AU on the pericentre. In general, the space-time geometry mainly tends to slightly reduce $p$, but drastically increases the eccentricity away from the apocentre, and thus, the semi-major axis as well (see the second and last columns, respectively). In contrast, the repulsive force ($\mathcal{FF}$) has the opposite effect and cancels out $\mathcal{ST}$ almost totally.

        Regarding the argument of pericentre (third column), the $\mathcal{ST}$ produces a characteristic feature of a negative slope around the apocentre, which is compensated for by the opposite feature produced by $\mathcal{FF}$. Near pericentre, the effects compete again, leading in the overall compliance with GR. Eq. \eqref{eq:delta_om_2pi} indicates that $\Delta \omega_{\text{MOG}} \simeq \Delta \omega_{GR}$ for this value of $\alpha$, which explains why we do not see a discrepancy between the solid blue and dashed black curve by eye. When we consider the findings from this figure, it becomes clear that the origin of the deviation between MOG and GR in Fig.~\ref{fig:orbital_elements} is the overcompensation of $\mathcal{FF}$ for the $\mathcal{ST}$ effect around the pericentre for high $\alpha$ values, while around the apocentre (grey shaded region), the contributions cancel out almost totally.

    \subsection{MOG versus DM}\label{sec:MOG_vs_DM}
        Considering the inspiration behind MOG, the interesting question arises whether the effect of a dark component on the stellar orbits is distinguishable from that of MOG. \cite{Heissel+22} showed that a DM profile demonstrates a characteristic feature on the evolution of the argument of pericentre, mainly in the range $f \in \left( 5\pi/6, 7\pi/6 \right)$, and that the overall precession $\Delta \omega_{DM}$ also decreases. We might therefore relax $M$ as a free variable and impose $\Delta \omega_{DM} = \Delta \omega_{GR}$ for a particular DM profile. Doing so for the MOG case as well, by relaxing $\alpha$ and $M$ and imposing $\Delta \omega_{\text{MOG}} = \Delta \omega_{GR}$, would lead to a direct comparison between the DM and MOG signature within an orbit.
        
\begin{figure}
            \centering
            \includegraphics[width=0.99\linewidth]{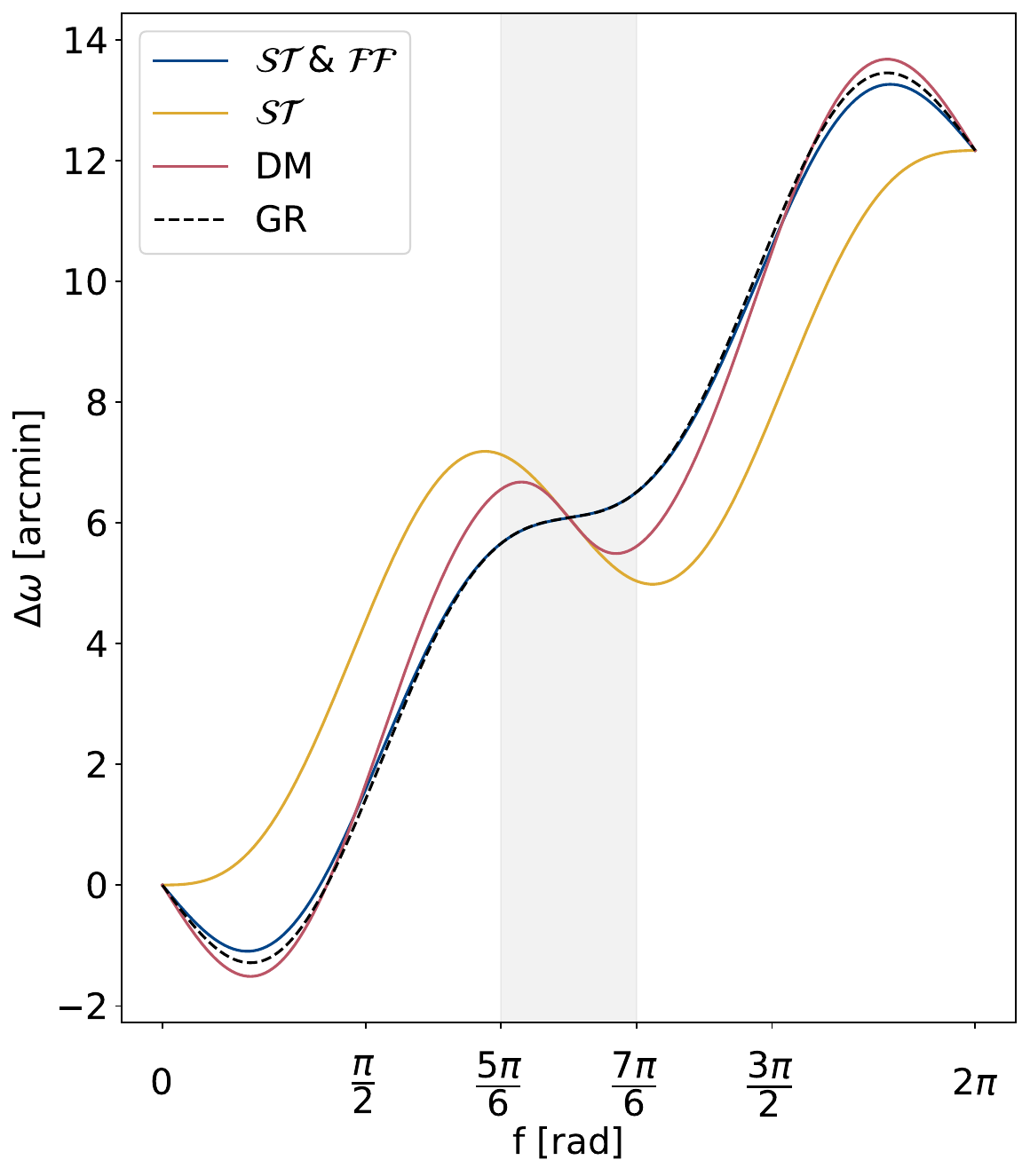}
            \caption{Evolution of the argument of pericentre with true anomaly. The full MOG theory is represented with a solid blue line, its $\mathcal{ST}$ effect with the solid yellow line, the DM case with the solid red line, and the GR case with the dashed black line. An explanation of the labels can be found in Table~\ref{tab:MOG_contributions}.}
            \label{fig:DMvsMOG}
        \end{figure}

        We refer to four cases, presented in Fig. \ref{fig:DMvsMOG}: i) full MOG ($\mathcal{ST}$ \& $\mathcal{FF}$), ii) $\mathcal{ST}$ contribution alone, iii) GR plus an extended mass (DM,) and iv) GR only. All the relativistic effects were taken into account up to 1PN. In the GR case, we selected $M = 4.261 \times 10^{6} M_\odot$, leading to the evolution of $\omega(f)$ represented with a dashed black line (Fig. \ref{fig:DMvsMOG}). Regarding MOG, we first took the full theory ($\mathcal{ST}$ \& $\mathcal{FF}$) for $\alpha =\alpha_\ast =\, 0.662$ \citep{Della_Monica_2021} and set $M = 2.205 M_\odot$, such that the overall precession matched (blue line).

        Regarding the extended mass, we considered the Plummer profile \citep{Plummer_1911},
        \begin{equation}
            \rho(r) = \rho_0\left(1 + \frac{r^2}{r_0^2}\right)^{-5/2},
        \end{equation}
        where $\rho_0$ is the density parameter, and and $r_0 = 0.3''$ is the length-scale parameter \citep{Mouawad_2005}. Following the most recent upper bounds on the extended mass \citep{GRAVITY+22_mass_distribution}, we chose the density parameter such that the total mass up to a radial distance of $230 \text{mas}$ was $3500 M_\odot$, corresponding roughly to $1\sigma$. To match the orbital precession, we selected $M = 5.43 \times 10^6 M_\odot$, leading to the red line in Fig. \ref{fig:DMvsMOG}. Finally, we isolated the $\mathcal{ST}$ effect (yellow line), and fixed $\alpha = 0.758 \times 10^{-3}$ and $M = 4.252 \times 10^6 M_\odot$, such that the slope at $f = \pi$ and the overall precession matched. All the other parameters were kept the same for all four cases.

        The full MOG theory, even for this high $\alpha = 0.662$ (which indicates an increase of 66.2\% in the gravitational strength), is almost identical to GR for almost the whole orbit. On the other hand, the DM case generates a characteristic feature around the apocentre, which was extensively discussed in \cite{Heissel+22}. It is significant that the $\mathcal{ST}$ effect achieves the same feature, although it deviates strongly far from the apocentre. In MOG, it is impossible for a massive test particle to follow a geodesic because of its $\mathcal{FF}$ charge $q_{TP} \sim m_{TP}$. However, when we take the relation between MOG BH and the Reissner-Nordström space-time (Sect.~\ref{sec:reiss_nord}) into account, it is evident that an electrically neutral test particle around a charged black hole in GR follows geodesics on a space-time that produces exactly the same feature. At very far distances compared to the BH scales, the $\mathcal{FF}$ weakens exponentially because the mass of the vector field is non-zero, and the $\mathcal{ST}$ effect therefore accounts for the missing mass. Therefore, these feature similarities should be expected.
        
        \begin{figure*}[b!]
            \centering
            \includegraphics[width=0.99\linewidth]{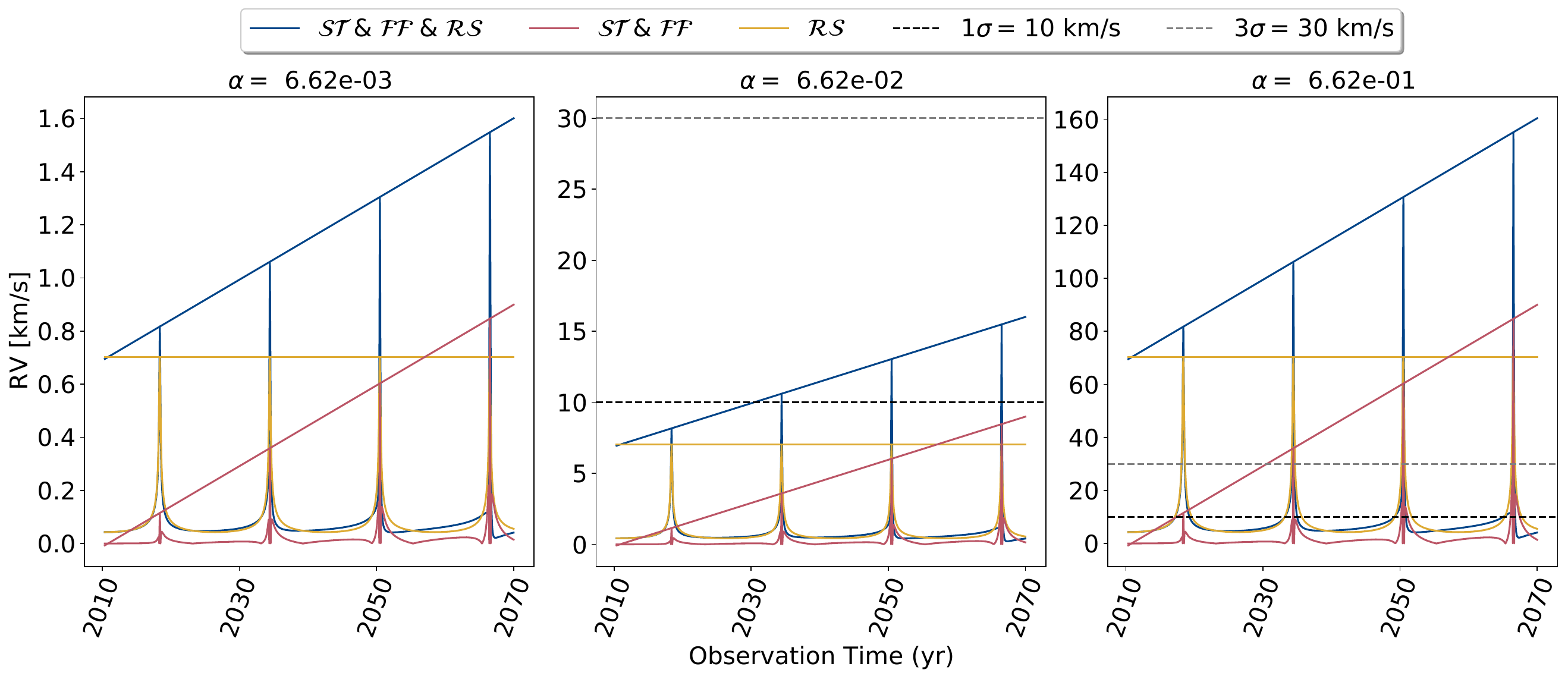}
            \caption{Evolution of RV deviation for three values of $\alpha$. The opaque lines (blue, red, and yellow, with spikes at the bottom) represent the RV absolute difference between i) MOG EoM with MOG $\mathcal{RS}$ and GR EoM with GR $\mathcal{RS}$ (blue), ii) MOG EoM without $\mathcal{RS}$ and GR EoM without $\mathcal{RS}$ (red), and iii) GR EoM with MOG $\mathcal{RS}$ and GR EoM with GR $\mathcal{RS}$ (yellow). An explanation of the labels can be found in Table \ref{tab:grav_red}. Instead, the transparent lines, namely, the progressive red and blue lines and the constant yellow line that connect the peaks of the respective opaque lines, represent the evolution of the corresponding RV absolute difference peaks. The horizontal dashed black and grey lines represent the 1$\sigma$ and 3$\sigma$ precision of the instrument, respectively.}
            \label{fig:redshift}
        \end{figure*}

    \section{Impact of redshift MOG versus GR}\label{sec:redshift}
        In all the investigations until now, we only used the equation of motion (EoM) up to 1PN (Eq.~\eqref{eq:MOG_eom}). However, for real data, the gravitational redshift ($\mathcal{RS}$) has to be taken into account as well. The spatial coordinates in Fig. \ref{fig:observables} are not affected by this phenomenon, unlike the radial velocity. Fig. \ref{fig:redshift} demonstrates three different comparisons to help us understand whether the gravitational redshift dominates the deviation of the local velocity due to the EoM. The comparisons are summarized in Table \ref{tab:grav_red}, where each of the two orbits (i.e. considering different configurations for EoM and $\mathcal{RS}$) are integrated separately and their resulting RV is then subtracted.
        \begin{table}[h]
            \renewcommand{\arraystretch}{1.2}
            \centering
            \caption{Orbits under investigation.}
            \begin{tabular}{|c|c||c|c||c|}
                \hline
                \multicolumn{2}{|c||}{\textbf{Orbit 1}} & \multicolumn{2}{c||}{\textbf{Orbit 2}} & \multirow{2}{*}{\textbf{Label}} \\
                \cline{1-4}
                EoM     & $\mathcal{RS}$    & EoM       & $\mathcal{RS}$    & \\
                \hline
                \hline
                MOG     & MOG               & GR        & GR    & $\mathcal{ST}$ \& $\mathcal{FF}$ \& $\mathcal{RS}$ \\
                MOG     & -                 & GR        & -     & $\mathcal{ST}$ \& $\mathcal{FF}$ \\
                GR      & MOG               & GR        & GR    & $\mathcal{RS}$\\
                \hline
            \end{tabular}
            \tablefoot{Orbits compared in Fig. \ref{fig:redshift}. Each orbit consists of the integrated EoM, and, eventually, the gravitational $\mathcal{RS}$ is taken into account. The second comparison ($\mathcal{ST}$ \& $\mathcal{FF}$) is also presented in the bottom row of Fig. \ref{fig:observables}, but on a logarithmic scale.}
            \label{tab:grav_red}
        \end{table}
        
        Fig. \ref{fig:redshift} shows that although the deviation between the MOG and GR EoM (i.e. $\mathcal{ST}$ and $\mathcal{FF}$) is progressive (red line), meaning that it increases with each period, the deviation due to the $\mathcal{RS}$ is constant with time. Initially, the $\mathcal{RS}$ has a stronger effect by one order of magnitude (yellow line), but the progressive deviation due to the EoM quickly kicks in and becomes dominant after some periods. Hence, for a sufficiently high $\alpha$ value and for a limited observation time, we have more opportunities to observe a discrepancy due to the $\mathcal{RS}$ (e.g. $\alpha = 0.662$), while for lower $\alpha$ values and for a long enough observation time, the EoM eventually produces a stronger deviation (e.g. $\alpha = 0.0662$). For very low $\alpha$ values, we can calculate how much observation time is required in order for the total deviation (blue line) to reach the instrument precision (dashed horizontal lines).

\section{Summary and discussion}
We have performed a detailed investigation of the first post-Newtonian effects of MOG on stellar orbits around BHs and compared the results with those predicted by GR. We found similarities between MOG and GR with electromagnetism, and we quantified MOG signatures on observables and orbital elements.

Namely, we confirmed the orbital precession ($\Delta\omega$) formula for MOG presented by \cite{Della_Monica_2021, 10.1093/mnras/stad579} using an alternative derivation (Eq~\eqref{eq:delta_om_2pi}). We further settled the disagreement that arose with \cite{10.1093/mnras/stac2113}, in Sect.~\ref{appendix:ffandem}.

We proved for the first time, to the best of our knowledge, an exact equivalence between dynamics around MOG and Reissner-Nordström BHs by proposing a fine-tuning (Eq.~\eqref{eq:tilde_identification} and~\eqref{eq:tilde_identification_test_particle}).

On the side of observable signatures, our analysis showed that for low values of $\alpha$, deviations from GR are minimal and difficult to detect within current precision. However, for higher values, the differences become significant, particularly in the orbital precession and in the evolution of orbital elements such as the semi-latus rectum and eccentricity. These effects manifest prominently in the pericentre half of the orbit and become more pronounced over multiple orbital periods.

By isolating the contributions of the MOG space-time geometry ($\mathcal{ST}$) and fifth force ($\mathcal{FF}$), we found that the two components can partially cancel each other out, leading to minimal deviations in the overall orbit. However, for high $\alpha$ values, the cancellation is incomplete, resulting in noticeable differences from GR predictions. This behaviour highlights the non-linear nature of MOG and the complex interplay between its components.

Additionally, we compared MOG to a DM scenario with an extended mass distribution. Interestingly, we found that certain MOG effects, particularly those related to space-time geometry, produce features similar to those seen by DM, such as the characteristic behaviour of the argument of pericentre around the apocentre. This suggests a potential degeneracy between MOG and DM, which might complicate efforts to distinguish between the two using current observational data.

Finally, we explored the effect of gravitational redshift ($\mathcal{RS}$) on radial velocity measurements. Our results indicate that while the redshift effect dominates initially, deviations due to the MOG EoM become significant over longer observation periods. This implies that precise and extended monitoring of stellar orbits is necessary to fully capture the MOG effect.

In conclusion, while MOG introduces subtle but potentially detectable deviations from GR, especially for higher values of $\alpha$, it remains challenging to distinguish between MOG and DM effects. Future observations, particularly those involving long-term monitoring of stellar orbits and high-precision measurements of radial velocity, will be crucial in determining the validity of MOG as an alternative to dark matter and in refining our understanding of gravity in the strong-field regime.

\begin{acknowledgements}
We thank N.~Aimar, T.~Paumard, G.~Perrin and F.~Vincent for fruitful discussions and access to the Galactic Centre orbit model code library of Paris Observatory/LESIA. This scientific paper was supported by the Onassis Foundation - Scholarship ID: F ZU 042/2 2025-2026. I. Liodis would also like to thank the Lilian Voudouri Foundation for their master's scholarship, the Onassis Foundation for their master's and doctoral scholarship, as well as Deutsches Elektronen-Synchrotron (DESY).
\end{acknowledgements}

\bibliographystyle{aa}
\bibliography{references}

\begin{appendix}

\section{Fifth force and electromagnetism}\label{appendix:ffandem}

We delve into a detailed analysis of the choice of sign in Eq. \eqref{eq:phu_mu_sign}, due to the confusion arose in the comment papers \cite{10.1093/mnras/stac2113} and \cite{10.1093/mnras/stad579}.

\cite{Lopez_Armengol_2017_b} studied the STVG-Kerr space time and invoked the resemblance of the vector potential $\phi_\mu$ around a black hole, where $\mu_\phi = 0$ with the electromagnetic four potential $A_\mu$ of the Einstein-Maxwell theory \citep{Misner:1973prb}. For a black hole with zero spin, the only non-zero covariant component of the vector potential is always negative, that is, 
\begin{equation}
    \phi_0 = - \frac{Q_{ff}}{r},
\end{equation}
where $Q_{ff}$ is the fifth force charge of the black hole and $r$ the Schwarzschild-like radial coordinate. Note that we now use natural units, since we only care about the direction of the extra force. It is important that \cite{Lopez_Armengol_2017_b} and \cite{Misner:1973prb} used the signature (-,+,+,+), from now on called $\mathcal{S}_{-}$. This is of course in accordance with electromagnetism in special relativity \cite{Srednicki:2007qs}, since the four potential around a point source of charge $Q_{em}$ is defined as
\begin{equation}\label{eq:contravariant_4potential}
    A^\mu = (\Phi, \mathbf{A}),
\end{equation}
where $\Phi = Q_{em}/r\,$,
leading to
\begin{equation}\label{eq:A3}
    A_0 = - \frac{Q_{em}}{r} {\quad \mbox{for } \mathcal{S}_{-}},
\end{equation}
due to the metric with $\mathcal{S}_{-}$ signature.

If one chooses the (+,-,-,-) signature (namely $\mathcal{S}_{+}$), the definition of the contravariant components of the four potential remains the same as in Eq. \eqref{eq:contravariant_4potential} \citep{Peskin:1995ev}, but now the covariant time component is given by
\begin{equation}\label{eq:A4}
    A_0  = \frac{Q_{em}}{r} {\quad \mbox{for }  \mathcal{S}_{+}},
\end{equation}
and the spatial components change sign. Note that in curved space-time, the covariant time components of the four potential have the same form with Eq. \eqref{eq:A3} or \eqref{eq:A4}, depending on the signature (see~\cite{Misner:1973prb}).

Going back to GR, it can be easily shown that in both cases, that is, using either $\mathcal{S}_{-}$ or $\mathcal{S}_{+}$, the physics is of course the same, obeying to the following equation of motion (see~\cite{Misner:1973prb,2016JTAP...10...47M}):
\begin{equation}\label{eq:EM_compact_eom}
    \dot{U} ^\mu +\Gamma ^{\mu }_{\alpha \beta }U^{\alpha } U^{\beta }=\frac{q_{em}}{m}F^{\mu }_{\phantom{a} \nu }U^{\nu},
\end{equation}
where $U^\mu$ is the four velocity of a particle with mass $m$ and charge $q_{em}$, moving in a space-time with Christoffel symbols $\Gamma ^{\mu }_{\alpha \beta }$ and in an electromagnetic field $F_{\mu \nu} = \partial_\mu A_\nu - \partial_\nu A_\mu$. Finally, the dot denotes derivative with respect to an affine parameter. For a point charge source placed at the origin, the only non zero components of the Faraday tensor are
\begin{equation}
    F_{1 0} = - F_{0 1} = \partial_r A_0 =
    \left\{
	\begin{array}{ll}
		+ Q_{em} r^{-2} & \mbox{for } \mathcal{S}_{-} \\
		- Q_{em} r^{-2} & \mbox{for } \mathcal{S}_{+}
	\end{array}
    \right.,
\end{equation}
while the radial component of the inverse metric can be expressed as
\begin{equation}
    g^{1 1} =
    \left\{
	\begin{array}{ll}
		+ \Psi_{\text{RN}}(r) & \mbox{for } \mathcal{S}_{-} \\
		- \Psi_{\text{RN}}(r) & \mbox{for } \mathcal{S}_{+}
	\end{array}
    \right.,
\end{equation}
where $\Psi_{\text{RN}}$ follows the Reissner–Nordstr\"om metric \citep{Reissner_1916,1918KNAB...20.1238N}. Therefore, the right hand side of the radial component of Eq. \eqref{eq:EM_compact_eom} is written as
\begin{equation}
     \frac{q_{em}}{m}g^{1 1} F_{1 0}U^{0} = + \frac{q_{em} Q_{em}}{m} \frac{\Psi(r)}{r^2} U^{0},
\end{equation}
considering $\mathcal{S}_{-}$ and $\mathcal{S}_{+}$. The only thing that defines the sign of this expression outside the horizon is therefore the product $q_{em} Q_{em}$, which is positive for like charges leading to repulsive force, and negative for unlike charges leading to attracting force.

The $\mathcal{FF}$ near a MOG BH has an equivalent picture if one identifies
\begin{equation}\label{eq:EM_MOG_identify}
\begin{array}{ll}
    F_{\mu \nu}         & \rightarrow \qquad B_{\mu \nu}, \\
    A_{\mu}             & \rightarrow \qquad \phi_{\mu},  \\
    Q_{em}              & \rightarrow \qquad Q_{ff},     \\
    q_{em}              & \rightarrow \qquad q_{ff},     \\
    \Psi_{\text{RN}}(r) & \rightarrow \qquad \Psi_{\text{MOG}}(r).
\end{array}
\end{equation}
However, now both charges are always positive \citep{Moffat_2015_BH}, and therefore, the $\mathcal{FF}$ is always repulsive, as desired.

\noindent
We highlight that the aforementioned correspondence leads to
\begin{equation}
    \phi_0 = + \frac{Q_{ff}}{r},
\end{equation}
for the signature that we used ($\mathcal{S}_{+}$), and that a different choice of sign would require either a non-physical absorption of the minus inside a charge, or  to postulate a relative minus sign in the action \eqref{eq:MOG_tp_initial_action}.

Retrospectively, it is evident that the claim of \cite{10.1093/mnras/stac2113} about the inconsistency of the signature used in \cite{Della_Monica_2021}, is correct. However, \cite{10.1093/mnras/stac2113} gave an expression for the orbital precession using $\mathcal{S}_{+}$ and a negative sign in the vector field, leading to an attracting force as demonstrated in this section, and as correctly pointed out in \cite{10.1093/mnras/stad579}. Finally, the latter study seems to have postulated a different sign in the $\mathcal{FF}$, in order to make it repulsive, which of course results in the correct equation of motion.

\section{Derivation of the 1PN Lagrangian}\label{sec:1PN_derivation}

The first term of Eq.~\eqref{eq:tp_action} can be obtained using the method followed by \cite{Gainutdinov_2020}, where the line element is given by \eqref{eq:MOG_line_element} and \eqref{eq:psi_MOG_1}. To comply with the notation in \cite{Gainutdinov_2020}, $\Psi (r)$ is written as
        \begin{equation}
            \Psi(r) = 1-2(1+\alpha)\frac{\mu}{r}+\alpha(1+\alpha)\frac{\mu^2}{r^2},
        \end{equation}
        with $\mu = G_N M/c^2$. One should now choose a proper coordinate transformation to go to isotropic coordinates with radial element $\rho$. Let the arbitrary transformation
        \begin{equation}
            r = \rho f(\rho), {\rm \ so \ that \ } \Psi^{-1}(\rho) \left(\frac{dr}{d\rho}\right)^2 = f^2(\rho).
        \end{equation}
        For positive $\alpha$, the differential equation obtained has the following solution
        \begin{equation}
        \label{eq:transformation}
            f(\rho) = 1 + \frac{(1+\alpha)\mu}{\rho} + \frac{(1+\alpha)\mu^2}{4\rho^2},
        \end{equation}
        and the line element is therefore written as
        \begin{equation}
            ds^2 = \Psi(\rho) c^2dt^2 -f^2(\rho) \left(d\rho^2 + \rho^2d\Omega ^2\right) \Rightarrow \nonumber
        \end{equation}
        \begin{equation}
            ds^2 = \Psi(|\mathbf{x}|) c^2dt^2 -f^2(|\mathbf{x}|) d\mathbf{x}^2,
        \end{equation}
        where $\mathbf{x}$ denotes the spatial position vector in Cartesian-like coordinates.
        Now, writing $\varphi_N = -G_N M/\rho = -c^2 \mu /|\mathbf{x}|$ and expanding $\Psi(|\mathbf{x}|)$ and $f^2(|\mathbf{x}|)$ with an accuracy of $\mathcal{O}(c^{-6} )$ and $\mathcal{O}(c^{-4})$, respectively, yields        \begin{subequations}\label{eq:Psi_f2}
        \begin{gather}
            \Psi(|\mathbf{x}|) = 1 + 2(\alpha + 1)\frac{\varphi_N(|\mathbf{x}|)}{c^2} + (\alpha + 1)(3\alpha + 2) \frac{\varphi^2_N(|\mathbf{x}|)}{c^4} + \mathcal{O}(c^{-6}),\\
            f^2(|\mathbf{x}|) = 1 - 2(\alpha + 1)\frac{\varphi_N(|\mathbf{x}|)}{c^2} + \mathcal{O}(c^{-4}).
        \end{gather}
        \end{subequations}
        For simplicity, from now on $\varphi_N(|\mathbf{x}|)$ will be referred to as just $\varphi_N$, implicitly having a dependence on $|\mathbf{x}|$. The line element is now written as
        \begin{align}
        \begin{split}
            \frac{1}{c} \frac{ds}{dt} =& 1 - \frac{\mathbf{\Dot{x}}^2}{2 c^2}
            \left[ 1 + \frac{\mathbf{\Dot{x}}^2}{4 c^2} - 3(\alpha+1) \frac{\varphi_N}{c^2}  \right] \\
            &+ (\alpha + 1)\frac{\varphi_N}{c^2}
            \left[ 1 + (2 \alpha + 1)\frac{\varphi_N}{2 c^2} \right] + \mathcal{O}(c^{-6})
        \end{split}
        \end{align}
        where the square root has been taken to an accuracy of $\mathcal{O}(c^{-6})$. Now, one can identify the geometrical term of the action \eqref{eq:tp_action} as $-c \,ds/dt$, leading to
        \begin{equation}
        \label{eq:lag1}
        \begin{split}
            -m c^2 \frac{d\tau}{dt} =& -mc^2 +m\frac{\mathbf{\Dot{x}}^2}{2}
            \left[ 1 + \frac{\mathbf{\Dot{x}}^2}{4 c^2} - 3(\alpha+1) \frac{\varphi_N}{c^2}  \right] \\
            &- m (\alpha + 1)\varphi_N
            \left[ 1 + (2 \alpha + 1)\frac{\varphi_N}{2 c^2} \right] + \mathcal{O}(c^{-4}).
        \end{split}
        \end{equation}
        Finally, in the isotropic coordinates, the $\mathcal{FF}$ term of the action \eqref{eq:tp_action} is written as
        \begin{equation}
        \label{eq:FF}
            - m c^2 \alpha \frac{G_N M}{r(\rho) c^2} = m \alpha \varphi_N + m \alpha(1+\alpha)\frac{\varphi_N^2}{c^2} + \mathcal{O} (c^{-4}).
        \end{equation}
        The full Lagrangian, including the $\mathcal{FF}$ \citep{Moffat_2009}, can be now read off from the action in the Cartesian-like coordinates, with the result given by Eq.~\eqref{eq:mog_lagrangian}.

\section{Auxiliary}\label{sec:aux_equations}
The transformation rules between the osculating orbital elements and position and velocity in Cartesian coordinates, at any instant of time, are given by:

\begin{equation}\label{eq:pos_and_vel}
   \qquad \mathbf r = (r^{X}, r^{Y}, r^{Z}) \qquad {\rm{and}} \qquad\mathbf v=( 
    v^{X}, v^{Y}, v^{Z})
\end{equation}
where
\begin{align*}
 & r^{X} = r\left( \cos\Omega\cos(\omega+f) - \cos\iota\sin\Omega\sin(\omega+f)\right) \\
 & r^{Y} = r\left(\sin\Omega\cos(\omega+f) + \cos\iota\cos\Omega\sin(\omega+f)\right) \\
 & r^{Z} = r\sin\iota\sin(\omega+f)
\end{align*}
and
\begin{align*}
& v^{X} =  -  C_p \big( \cos\Omega(\sin(\omega+f) + e\sin\omega)\\
&\phantom{v^{X}} = C_p( + \cos\iota\sin\Omega(\cos(\omega+f) + e\cos\omega\big)\\
& v^{Y} =-C_p \big(\sin\Omega(\sin(\omega+f) + e\sin\omega) \\
&\phantom{v^{Y}} = (- \cos\iota\cos\Omega(\cos(\omega+f) + e\cos\omega\big)\\
& v^{Z} = C_p \sin\iota(\cos(\omega+f)+e\cos\omega)\, ,
\end{align*}
with $r =p/(1+e \cos f)$ and $C_p =  \sqrt{Gm/p}\,$.
For their derivation  see, for instance,~\cite[Sect.~3.2.5]{PoissonWill2014}.
\end{appendix}

\end{document}